\documentclass[12pt,preprint]{aastex}

\received{}
\accepted{}

\slugcomment{}

\shortauthors{Aller et al.}
\shorttitle{ Properties of the Pearson-Readhead Sources}

\begin{document}

\title{Pearson-Readhead Survey Sources II: The Longterm Centimeter-band 
Total Flux and \\
Linear Polarization Properties of a Complete Radio Sample}

\author{M. F. Aller, H. D. Aller, and P. A. Hughes}
\affil{Department of Astronomy, University of Michigan, 817 Dennison Building, Ann Arbor, MI 48109-1090}
\email{margo,hugh,hughes@astro.lsa.umich.edu}

\begin{abstract}
Using UMRAO centimeter-band total flux density and linear polarization
monitoring observations of the complete Pearson-Readhead extragalactic
source sample obtained between August 1984 and
March 2001, we identify the range of variability 
in extragalactic objects as functions of optical
and radio morphological classification, and relate total flux
density variations to structural changes in published coeval VLBI maps in 
selected objects. As expected, variability is common in flat or inverted
spectrum ($|\alpha|\leq 0.5$) core-dominated QSOs and BL Lacs. Unexpectedly,
we find flux variations in several steep-spectrum sample members, including
the commonly adopted flux standard 3C~147. Such variations are characteristically
several-year rises or declines, or infrequent outbursts, requiring long-term 
observations for detection: we attribute them to the brightening of weak core
components, a change which is suppressed by contributions from extended structure in all
but the strongest events, and identify a wavelength-dependence for the
amplitude of this variability consistent with the presence of
opacity in some portions of the jet flow. One morphological class
of steep-spectrum objects, the compact symmetric objects, 
characteristically shows only low-level 
variability. We examine the statistical relation between
fractional polarization and radio class based on the data at 14.5 and
4.8 GHz. The blazars typically exhibit flat to inverted
polarization spectra, a behavior attributed to opacity effects. Among
the steep spectrum objects, the lobe-dominated FR Is have steep fractional
polarization spectra; while the FR IIs exhibit fractional polarization
spectra ranging from inverted to steep, with no identifiable 
common property which accounts for the range in behavior.
For the cso/GHz-peaked-spectrum sources we verify that the fractional
polarizations at 4.8 GHz are only of order a few tenths of a percent; but
at 14.5 GHz we find significantly higher polarizations, ranging from one
to three percent; this frequency dependence supports a
scenario invoking Faraday depolarization by a circumnuclear torus.
We have identified preferred orientations of the electric vector of the
polarized emission at 14.5 and 4.8 GHz in roughly half of the objects,
and compared these with orientations of the flow direction indicated by VLBI
morphology. When comparing the distributions of the orientation offsets for
the BL Lacs and for the QSOs, we find differences in both range and mean value, 
in support of intrinsic class differences. In the shock-in-jet
scenario, we attribute this to the allowed range of obliquities of shocks
developing in the flow relative to the flow direction: in the BL Lacs
the shocks are nearly transverse to the flow direction  while in the QSOs they
include a broader range of obliquities and can be at large angles to it. The 
fact that we find longterm stability in EVPA over many events implies that
a dominant magnetic field orientation persists; in the core-dominated objects, 
with small contribution from the underlying quiescent jet, this 
plausibly suggests that the magnetic field has a longterm memory, with
subsequent shock events exhibiting similar EVPA orientation, or, alternatively, 
the presence of a standing shock in the core. We have
looked for systematic, monotonic changes in EVPA  which might be expected
in the emission from a precessing jet, a model currently invoked for some AGNs;
none were identified. Further, we carried out a Scargle periodogram analysis
of the total flux density observations, but found no strong evidence for periodicity 
in any of the sample sources. The only well-established case in support of both jet
precession and periodic variability remains the non-sample member OJ~287.

\end{abstract}

\keywords{extragalactic objects: general--- galaxies: active --- polarization --- quasars: general --- shock waves}

\section{Introduction}
In August 1984 we initiated a program to systematically monitor the sources
in the flux and position-limited Pearson-Readhead  \citep{per81,per88}
VLBI survey at cm wavelengths using the University of Michigan 26-meter paraboloid
operating at 14.5, 8.0, and 4.8 GHz, with the primary goal of delineating the range 
of behavior in the total flux and linear polarization in a complete flux-limited
sample of extragalactic objects. The source group includes representatives of
the optical classes QSO, galaxy, and BL Lac object, as well as of several VLBI 
morphological classes, making it ideally suited  for exploring variability properties
as a function of source type. The limit on source flux density provided by the defining 
criteria for the sample ($S\ge 1.3$ Jy at 5~GHz) ensured that our observations
would provide high signal-to-noise polarization measurements for most sources,
since the linearly polarized flux from extragalactic objects is typically on the
order of a few percent \citep{all96}.

Results based on data obtained through 1991.0, and including some
heterogeneous  earlier measurements for well-known variable sources, were
presented in \citet{all92} (hereafter Paper I), which
discussed the statistics of the flux variability and polarization and
related them to  optical classification and to VLBI morphological
classification taken from the literature. The VLBI data then available
for comparison were primarily at 5~GHz and lower frequencies. Paper I 
demonstrated that the  QSO group was comprised of both flat and steep spectrum objects
with a skew towards flat spectra and that most QSO members were variable, 
with higher amplitude
variations present at 14.5~GHz than at 4.8 GHz, while the galaxies in
general exhibited steep spectra, lower-amplitude-to-no flux variability,
and longer-term stability in the orientation of the electric vector of the
polarized emission (EVPA). This was attributed to the fact that the major
contributor to the galaxies' integrated radio emission in most sources was
dominated by a contribution from the underlying extended structure 
rather than from evolving, compact, core (and jet) 
components. 

In the current paper we reexamine those source properties which can be investigated
through multifrequency single-dish monitoring data which more than double the time
base of Paper I. Our measurements are now
supplemented by a wealth of published complementary radio
interferometric data at both centimeter and millimeter wavelengths,
which has become available through large survey studies
undertaken with the VLBA. These include multi-epoch total intensity images
matching our own spectral band \citep{kel98}, as well as
total intensity and polarization imaging at a higher frequency, 43 GHz, 
\citep{lis01a,jor01} probing deeper into the inner jet regions. In
combination, the single-dish and imaging data provide additional insights
into the emission properties of extragalactic objects.
 
\section{OBSERVATIONS}

\subsection{The Sample}

The Pearson-Readhead sample \citep{per88} (hereafter PR) uses the following selection criteria:
1) declination (1950) $\delta>35^{\circ}$; 2) galactic latitude
$|b|>10^{\circ}$; and 3) total flux density S$\geq$ 1.3 Jy based
on the NRAO-MPIfR 6 cm strong source surveys.
Of the sixty-five sources in the sample, sixty-two are within the
declination limit, $\delta<82^{\circ}$,
of the University of Michigan 26-m paraboloid. These sources 
have been systematically observed once every three months at 14.5, 8.0 and 14.5~GHz
since fall 1984, following the observing technique and reduction procedures
described in \citet{all85} and  \citet{all03}, with scaling
to conform to the flux calibration system of
\citet{ott94}. In addition to the trimonthly observations, sources identified
as exhibiting high degrees of activity were incorporated into our core
program for more frequent (approximately  weekly) observation. Three objects,
DA~55, 1928+738, and BL Lac, are also part of a VLBA study we are undertaking
to follow the structural evolution at centimeter and millimeter wavelengths
in highly active objects \citep{all01}, and hence the data sampling for these
sources during the past few years is particularly frequent.

\subsection{Emission Properties}

The program sources and emission parameters are given in Table 1.
These include optical ID, redshift, the radio class based on
morphology, both a variability index and a
fluctuation index which are discussed below, and the spectral index between
14.5 and 4.8 GHz. 

The optical class listed in column 3 for each object is
taken from the literature; the sample
contains 29 QSOs (Q), 25 galaxies (G) , and 8 BL Lac objects (BL). Of the 8 BL Lacs,
6 were also included in the UMRAO flux-limited 
BL Lac sample \citep{all99}; the remaining two objects, 0814+425
and 1823+568, did not met the inclusion criteria for the BL Lac sample
and are not listed in \citet{bur87},
but are included in the broader compilation given in \citet{bur92}. We note that 
the optical classifications
for three objects have been revised since Paper I as follows: 0212+735 (BL to Q),
3C 371 (G to BL) and OW 637 (G to Q). For 0710+437 we have retained the
class G, given in \citet{ode98} and \citet{pec00}, which differs from the
class given in NED. 

A diversity of VLBI morphological type, as shown in column 
4, is also represented by the member sources. These include
compact (c), lobe-dominated (ld),
asymmetric (a1 and a2), irregular (i), steep
spectrum compact (ssc), following the classification scheme of
\citet{per88}, and  compact symmetric (cso). By radio morphology
five steep spectrum compact (ssc)  objects and seven confirmed or
suspected compact 
symmetric objects (cso), members of the COINS sample \citep{pec00} (identified
or candidate compact symmetric objects in the northern sky) are included. A subset
of five of these COINS objects (0108+388, 0710+439, 1031+567, 1358+624, \& 2352+495)
are members of the \citet{stan98} 1~Jy sample of GHz-peaked-spectrum
sources (hereafter GPS), 
a group characterized by convex radio spectra with observed turnovers (uncorrected
for redshift) near 1 GHz. The  centimeter-band spectra of the GHz-peaked sources
are typically steep in the high frequency, presumably transparent,
part of the spectrum, and past work has characterized
them as exhibiting low degrees of variability and low fractional linear
polarization \citep{stan98}. These objects are particularly
intriguing because they are  postulated to be young on the basis of their small
intrinsic sizes, although it has been alternatively proposed that they are small
due to confinement by an ambient medium \citep{ode98}. Four of these GPS sources 
have been monitored with the VLBA at 15 and 43 GHz to determine motions and
inferred kinematical ages \citep{tay00b}, yielding ages of 300-1200 years for the 
current radio activity. Column 5 lists 
redshifts predominantly taken from NED and from the latest version of the Veron
\& Veron catalogue available on the web (http://www.obs-hp.fr/). Column 6 
gives the same variability index, V, used in Paper I at 14.5 GHz and defined
to be:

\begin{equation}
V=\frac{(S_{max}-\sigma_{Smax})-(S_{min}+\sigma_{Smin})}{(S_{max}-\sigma_{Smax})+(S_{min}+\sigma_{Smin})}
\end{equation}
This variability index is a measure of the peak-to-trough 
amplitude change in the total flux during the time window of the observation.
In computing parameters characterizing source emission, observations
with $\sigma_{S}>$max(0.1Jy,  0.03S) were excluded to remove observations
with unusually poor signal-to-noise. Additionally, some
measurements were rejected because of source proximity to the sun. 
At 14.5 GHz the number of accepted daily averages searched to identify the
minimum and maximum fluxes ranged from 53 to 766, with 
a median value of 85.

While V is an appropriate measure of the flux amplitude change in variable sources, 
it may not be a useful measure of variability in sources with low-amplitude variability
(V$\leq$ 0.2) where the variations approach the signal-to-noise level of the
measurements. Such values are rare at 14.5 GHz, but are more common at 4.8
GHz where the amplitude of the variability is generally reduced. (See Paper
I, Table 1). Also, other investigations have used a measure related to the
standard deviation of the data, especially when attempting to identify
low-level variability. Therefore, in columns 7 and 8 we include a second variability
measure, a fluctuation index  which measures the spread about the mean flux and defined as:
 
\begin{equation}
FI= \left( \left({{\sum\nolimits X_i^2\omega_i-\overline{X}\sum\nolimits X_i\omega_i} \over {N-1}}-1 \right) \cdot {N \over {\sum\nolimits \omega_i}} \right)^{0.5} \cdot {1 \over \overline{X}}
\end{equation}
Here N is the number of observations of the variable $X_i$ with measurement
uncertainty $\sigma_i$ and weight $\omega_i = \sigma_i^{-2}$. 
Note that this index attempts to remove the
scatter due to measurement errors: an infinite number of measurements of a
source where all the scatter is due to normally distributed measurement
errors would yield an expected value of zero.
In Figure 1 we show a plot of V versus FI, illustrating that these two
indices are highly correlated and showing that V is a reliable measure 
of variability, permitting comparison of the new variability results
with those of Paper I.

Column 9 lists average spectral indices using data 
at 14.5 and 4.8 GHz. These were computed from paired values of
monthly averages; this procedure 
ensured a time match between the two frequencies for the most poorly
observed sources. We adopt the sign convention that $S\propto\nu^{+\alpha}$. 
Column 10 indicates sources included
in higher-frequency VLBI surveys: \citet{lis01a,jor01,kel98} coded L, J, and K
respectively. Information on
sources which were added to the NRAO 2 cm survey subsequent to \cite{kel98}, and
which are available on the World Wide Web via the NRAO URL, are also included.

\placetable{tbl-1}

\section{TOTAL FLUX DENSITY VARIABILITY}

\subsection{Variability Amplitude as a Function of Optical Class and Redshift}

Compared to our results on variability from Paper I, the most striking 
differences using the extended time base occur for galaxies: 
while 11 sources previously showed no statistically significant variation
at 14.5 GHz, several of these now show variability, as 
shown in Figure 2 and discussed below. In contrast to the outbursts seen in AGNs,
the behavior of the variability in the galaxies is characterized by
slow changes, either longterm events (e.g., 3C 84 which has shown 
a single rise and fall extending
over nearly four decades), slow, nearly monotonic changes, or 
infrequent, discrete events (e.g. 3C 390.3: Figure 11).
These behaviors are described in detail in section 3.4.

The types of objects showing  the largest amplitude variations are
BL Lacs and QSOs. A KS test of the distribution of variability
index V for the BL Lac objects and for the QSOs gives a probability
of 59$\%$  that these distributions are drawn from the
same population. However, even among
the small number of BL Lacs in the
sample, there are differences in the details of the variability. 
These range from large
amplitude variations in 0954+658  to lower-amplitude
fluctuations in Mkn 501; in the latter a structure function analysis 
of our early data \citep{hug92} identified that
white noise, rather than shot noise typical of AGNs, characterized 
the emission process (see also section 4.5).

A relation between variability and redshift might be expected. Therefore,
in Figure 2 we plot the fluctuation index at 14.5~GHz versus redshift
as a function of optical class. While the redshifts for the galaxies (and
BL Lacs) fall in a narrower range (z$<$1.0) than for the QSOs, we find
no obvious relation between FI and redshift for the sample as a whole. 
That the amplitude ranges are the same for the BL Lacs
and QSOs is clearly evident in this figure.

\placefigure{fig2}

\subsection{Time-Averaged Total Flux Spectrum; Relation to Optical Class and Radio Morphology}

Radio spectral index is used as a class delineator in the analysis that
follows. To assess whether there are changes in the time-averaged
spectra depending on the duration of the time window, we have compared
the average spectral indices
shown in Table 1 with those determined using data from the first half of the
time window only. We find no difference in average $\alpha$ within the
standard deviation for each source, and conclude that our average spectral
indices appropriately represent the centimeter-band spectra.

Among the sources classed as galaxies, we find only one flat spectrum  
($\vert\alpha\vert\leq 0.5$) object, 3C~84, a well-known variable source
which has exhibited large-amplitude, slow, total flux variations. The
remainder of the galaxies have steep spectra ($\vert\alpha\vert>0.5$). In
spite of steep total flux spectra, we find that nine of the twenty-five steep-spectrum
galaxies vary in flux density at 14.5~GHz with FI$\geq0.08$ and V$\geq0.17$. 

The majority of QSOs have flat spectra, and are variable. However, seven of
the twenty-nine QSOs do have steep spectra: 0153+744, 0538+498 (3C 147), 0711+356, 
0809+483 (3C~196), 1458+718 (3C~309.1), 1634+628 (3C~343), and 1828+487 
(3C 380). By radio morphology, this
group includes two compact doubles, four steep spectrum compact objects,
and one lobe-dominated source. All of
the BL Lacs have flat time-averaged spectra. 

  As illustrated in Figure 3, at 14.5 GHz most steep spectrum and some 
flat spectrum objects are found to 
be variable, while at 4.8 GHz there is a sharp demarcation 
between the fluctuation indices for these two groups.
The distribution of fluctuation index with alpha may be explained in the context
of unification schemes invoking orientation-dependent beaming effects. The curves shown
superimposed on the 14.5 GHz data in Figure 3 result from computations of a
two-component model: an extended, non-variable (presumably unbeamed)
component with a spectral index of -1.0, and a time variable compact 
core with a spectral index of 0.5. The locus arises from varying
the ratio of the flux density of the core and extended component (as would
be expected if a beamed core were observed over a range of viewing angles).
The solid curve shows the locus for a core of fixed fluctuation 
index of $0.5$, and the dashed curve shows the case where the core's fluctuation 
index increases as the observed core flux increases. Such a boost in 
fluctuation index would result, for example, if the peak flux were preferentially
boosted relative to the minimum `quiescent' flux for cores seen with the flow
axis close to the line of sight, but would be little boosted for edge-on flows.
Such a scenario would apply if flow speed is higher during outbursts, or if
outbursts reflect distinct emission regions boosted by an additional power of
the Doppler factor: in the absence of a particular model, we adopt a linear
relation between fluctuation index and core flux as an example of possible
behavior. Even a simple proportionality of the fluctuation index to the Doppler boost 
factor in the modeling gives a clearly better fit to the observed distribution.
The low level variability in galaxies probably occurs because 
in a flux-limited sample they were preferentially selected 
by their strong extended components rather than by active cores 
beamed toward us.

\placefigure{fig3}

\subsection{Effects of Bias on Sample Selection}

The effects of beaming and redshift on duty cycle, and hence on 
source selection, have been investigated by \citet{lis01b} 
for flat-spectrum AGNs; he concludes that variability
will generally be unimportant in the selection statistics, at least for
sources whose flux amplitudes place them well above the threshold value.
We concur with \citet{lis01b} that this sample
is relatively free of bias in source selection, and we
believe that it is an appropriate representation of extragalactic
objects for statistical study.
We note, however, that the full PR sample does, in fact, contain
non-blazar-like but variable objects, in addition to presumably highly
beamed sources exhibiting blazar-like variability. Additionally, several
sources {\it do} have fluxes near the survey limit; as shown in Figure 4, 
the time-averaged flux at 4.8 GHz for 1/4
of the sources lies at or below the flux limit for sample inclusion
during the period of our study, and one-third of the sample objects
decreased in flux to amplitudes below the flux cutoff for durations
of one to several years:  0711+356, has
remained below the original 5~GHz limiting flux during the entire
duration of our study. The possible influences of these factors have not
been explored quantitatively.

\placefigure{fig4}

\subsection{Spectral Evolution and Variability Changes}

While statistically the values of the time-averaged spectral indices 
used to separate the flat and steep spectrum objects have not changed
on time scales of several years, 
we identify significant changes in the variability properties 
of several individual objects  We present and discuss below the light curves
for selected objects to illustrate the range of variability we identify.
Monthly averages of the UMRAO flux and polarization for all sample  members will be
included in \citet{all03}. We show as examples:

\subsubsection{a flat-spectrum QSO}

  In Figure 5 we show the flux and polarization light curves for the
QSO 1637+574. The behavior is typical of our program AGNs with large
amplitude changes (V(14.5)=0.54) 
and a  flat to inverted spectrum ($\alpha_{av}=0.16$). The April 1999
43 GHz VLBA map of \citet{lis01a} shows the source to consist of a core
and two jet components with a `bend-and-realign' morphology attributed
to flow along a helical path. A large event
in the mid-1990s is the dominant feature in the total
flux density light curve. Note that the
solid line in the lower panel marks the 5~GHz total flux limit for
inclusion in the sample. During the 2 year period
1997-1999, and for shorter earlier periods, the total flux at 4.8 GHz
was below the flux criterion for sample inclusion.

\placefigure{fig5}

\subsubsection{an ssc QSO with a masked variable core component}

In Figure 6 we show yearly averages of our longterm data for the QSO
3C~147 which is a prototypical steep spectrum compact source and
a `typical' quiescent source proposed as a secondary flux standard 
\citep{baa77,ott94}. Lister's VLBA map
\citep{lis01a} shows a core and some amorphous extended structure at
43 GHz, but no well-defined jet components and a total image size of
4 mas. At 5~GHz, VLBI imaging reveals a jet extending 200 mas to the
southwest. The UMRAO integrated total flux data show a steep (transparent)
total flux spectrum ($\alpha_{av}=-0.98$) as expected. 
However, while there are no dominant outburst-like events in the total
flux light curve, small systematic changes in the flux are evident on
a timescale of several years. These are
more evident in Figure 7 which displays yearly averages
of the total flux density, with a flux corresponding to the spectrum
given by \citet{ott94} subtracted from each measured UMRAO datum.
Removing this assumed constant component, based on observations
obtained during 1989-1992,
leaves a clearly identifiable variable component, illustrating that
this is a case where the variability of the source 
is masked by the dominant contribution from extended structure. 
The fact that the source is variable can be seen directly in
the integrated polarization
(middle and top panels of Figure 6). Note that the polarized flux shows
an inverted spectrum: the source is measurably  but only
weakly polarized at 4.8 GHz but exhibits stronger polarized flux at
14.5~GHz, at a level comparable to that typical of AGNs \citep{all99}.
The  EVPAs show a pronounced frequency-dependent separation consistent with,
but not totally explained by, an average local rotation  measure  of
-1300 rad m$^{-2}$ reported for the core (Nan et al. 2000). (Complex
 polarization structure has  been identified from VLA
mapping \citep{jun99}, so we would not expect an unambiguous signature
of Faraday rotation in our data.) Based on our data we cannot
assess the competing importance of synchrotron self-absorption and
Faraday depolarization near the core.

\placefigure{fig6}
\placefigure{fig7}

\subsubsection{an ssc QSO exhibiting resolved variability}

Figure 8 shows the total flux density and polarization light curves
for a second ssc source, QSO 3C 309.1. Unlike the behavior seen in 3C 147,
the EVPAs  do not show a frequency-dependent offset; and rotation measure
mapping has revealed typical rotation measures of only 60 rad m$^{-2}$, with
smaller values in the lobe \citep{aar98}. Thus, the physical conditions are
different for the two sources in spite of their belonging to the same optical and
radio types. Prior to 1992, the total flux light curve is similar to that 
for 3C~147, exhibiting only low level variations. However, subsequently
there is a  remarkable change characterized
by a large total flux density increase at both 8 and 14.5~GHz, and a
corresponding change in spectral index. Significant changes both in the
centimeter-band spectrum and in the amplitude of the variability can occur
with time scales of order a decade in members of this radio class.

\placefigure{fig8}

 A similar, but less dramatic flux event than that shown in Figure 8
has been seen in a third ssc source in the sample: 3C~380. While it might
be expected that ssc sources would not exhibit blazar-like activity, it has
been proposed that there are in fact two classes of ssc QSOs, those with
prominent cores and superluminal behavior such as
3C~380, 3C~147, and 3C~390.1, and those with weak cores
\citep{aku92}. Our data show evidence for blazar-like activity
in both 3C~380 and 3C~390.1 which must be associated with changes
in evolving components and which would have been missed by a shorter 
time window because of the length of the times between
quiescent and active periods. Indeed, our impression of the
variability in 3C~390.1 based on data through 1991.0 was very
different from that including the later data. A fourth object from 
our group with steep centimeter-band spectra, the compact double
QSO 0153+744, shows a longterm monotonic decrease in total flux at both
4.8 and 14.5~GHz.

\subsubsection{a cso object exhibiting low amplitude flux variability typical of the class}

In Figure 9 we show yearly averages of the UMRAO data for the 
galaxy 0108+388, classed as compact symmetric based on its radio
morphology and believed to be one of the youngest members of this class
based on its kinematical age of 367 years \citep{pol02}. Because the polarized 
flux is weak, several EVPAs have not met our acceptance criteria
(P/$\sigma_{P}>2$) for plotting, and hence are not shown. 
The source is a member of the Stangellini 1 Jy GHz-peaked source sample, 
and we find that it exhibits both low degrees of total flux variability
(FI(14.5)=0.09), and a low degree of average fractional polarization at 5~GHz  
($P_{av}=0.30\pm$0.07\%), as expected for GPS class members.
High resolution VLBI images document that the structure of this source is
complex, and 15~GHz VLBA observations of total intensity have been modeled
assuming 7 components, with the largest fraction of the flux contained in the
outermost components on either side of the core, and a relatively low
flux contained in the central component identified as the core and center of
activity \citep{tay96}. The very low level flux variability seen in the bottom
panel is typical of csos. Our longterm data support the proposal by
\citet{fass01} that objects of this class be used as VLA calibrators
because of the relative stability of their flux levels.

\placefigure{fig9}

\subsubsection{a possibly misidentified cso: variable!}

In Figure 10 we show the data for a tentatively identified compact 
symmetric object (cso) 2021+614 (OW~637). While exhibiting
unusual spectral behavior with a turnover at 8.4 GHz, the source is not
included in the 1 Jy  GPS survey
\citep{stan98}, nor included in O'Dea's comprehensive list of 
css and GPS sources \citep{ode98}. The series of maps from 1994 through 2001 
for this source, obtained as part of the 2cm VLBA survey, show the source to be
double-sided \citep{kel98}, as does the 43 GHz map of
\citet{lis01a}. Our light curves show a time-variable
spectrum, significant variability at 14.5 and 8 GHz, and a longterm
rise at 4.8 GHz. The low degrees of polarization we find are typical
of those for GPS sources, but the character of the spectrum and the flux
variability are time dependent, very different from those found
for 0108+388, and not characteristic of a GPS source. The variability at 
14.5~GHz is, in fact, similar to that seen in our program blazars. Lister 
finds only upper limits on the fractional polarization for the components
he has identified at 43 GHz, a result consistent with the low fractional
polarization which we find in the centimeter band. A subluminal pattern 
speed of $0.12\pm0.02h^{-1}c$ has been determined by \citet{tsc00}. On the
basis of the amplitude and timescale of the variations we find, we would
have expected superluminal motion.

\placefigure{fig10}

\subsubsection{an ld (lobe-dominated) galaxy with a variable component}

Figure 11 shows the flux and polarization for the X-ray bright,
lobe-dominated, FR II radio galaxy 1845+797 (3C 390.3). Our data through 
1991.0 included in the analysis for Paper I showed some evidence of flux 
variability, but the data were undersampled for following these
variations in detail. The sampling was subsequently increased at 14.5~GHz, and
these data reveal a well-defined, monotonic swing in EVPA at 14.5~GHz during 
very late 1994-early 1995 associated with the total and polarized flux event 
which occurred during this period. This result is consistent with  the VLBI
results of \citet{pre96} who find a strong jet feature in their 1995.29 map 
at 5~GHz which was much weaker in a 1989 image.

\placefigure{fig11}

 \subsection{Summary of Variability Properties}
 We conclude that variability is not restricted to flat spectrum
sources and that its characteristic behavior in an individual source
can change on time scales of order a decade. Such changes in
behavior have previously been
identified in UMRAO observations of a few non-sample members, e.g. 3C~120,
and OJ~287 \citep{all96}, both flat spectrum, core-dominated AGNs. 
Multi-epoch VLBI observations demonstrate
that changes in these objects are consistent with the development and
evolution of hotspots in the flows of QSO and BL Lac jets. Variability
in galaxies is most often not outburst-like, as is commonly found in BL Lacs
and QSOs, but rather consists of slower variations, often only identifiable
with a time window of order a few decades, or infrequent events,
also requiring a long time window to detect variations in the integrated
total flux data. We attribute the behavior we see in galaxies
to relatively long timescales combined with the low luminosity of the active
core components relative to that from the extended structure, requiring
an unusually bright event for detection.

\section{LINEAR POLARIZATION}

Long term linear polarization observations provide information on
the characteristics of the quiescent flow - namely the degree of turbulence
and the dominant orientation of the magnetic field in the radio-emitting region
since, in
the optically thin case, the magnetic field orientation is orthogonal to the 
directly-measured EVPA. Detailed modeling of selected time segments of our
multifrequency total flux density and linear polarization observations in three
sources exhibiting the expected signature of a propagating shock has provided
strong evidence for the passage of a transverse
shock in these selected cases \citep{hug91}. In other objects, comparison of 
changes in the EVPA  during active phases and the 
VLBI parsec-scale structural axis, a measure of the flow direction in the jet, 
has demonstrated that shocks are most often at oblique angles to the flow
direction rather than transverse to it. Such comparisons
have used both integrated polarization observations, when a single component is
dominant, \citep{all98} and VLB polarization imaging \citep{all00,lis01a,all02}
in cases of complex, as well as simple, structure.

Table 2 contains values of the preferred orientation of the electric
vector of the polarized emission and the flow direction indicated by
VLBI morphology. Columns 2 and 3 give the adopted rotation measure taken
from the literature and its reference. Note that in some well-studied
AGNs (of which a few are included here), the central regions have
been found to have very large rotation measures
which differ from the relatively low rotation measures in other parts of
the radio source  \citep {tay00a}, and which are quite different
from the integrated values adopted here. In columns 4 and 5 we give the value
of the average percentage polarization at 14.5 GHz computed using the Stokes
parameters for the daily observations, and the associated rms scatter
(a combination of the measurement error and the variability of the source);
these tabulated polarizations have been corrected to avoid systematic bias
when polarization measurements with different signal-to-noise
ratios are combined \citep{war74}. The EVPAs in columns 6 and 8 were obtained
following the procedure described in detail in \citet{all99}. 
For each source and frequency, a histogram of EVPA was constructed based on
monthly averages of the Stokes parameters after correcting for Faraday rotation, 
using the rotation measures
given in column 2. Using a chi-squared test, statistically significant
peaks in each distribution were identified which met the
acceptance criterion of $\chi^2/N\geq$3; the fraction of time
that the preferred orientation was maintained was also computed.
These fractional times are shown in columns 7 and 9. 
 In columns 10-15 we tabulate information on the VLBI parsec scale 
source orientations compiled from the literature at both centimeter and
millimeter wavelengths, and including values of $\theta$ measured by us
from published maps when no values were cited. 
The suffixes u, b, and c denote unresolved structure, small scale bends,
and largescale curvature respectively, based on our impression of
the source structure in the VLBI maps. It is well-documented that there are commonly
significant single, or multiple, bends in the jets of AGNs, and in values measured
by us we have weighted the dominant parsec scale flow direction $\theta_{5,15,43GHz}$
towards the innermost, best-defined components, or selected from the literature
the values based on the inner structure. While
maps are available for the compact symmetric objects, their morphology is
unusually complex, in several cases exhibiting an `S' symmetry,
possibly due to precession of the central engine \citep{tay96}, and
hence there is not a single, well-defined parsec scale flow direction
for these sources. Except for 0404+768, our data show no preferred
EVPA at either 4.8 or at 14.5~GHz  for members of this group. 

\placetable{tbl-2}

\subsection{Range of Linear Polarization as a Function of Radio Class}

We have examined the average fractional polarization as a function of radio 
and optical class. We show in Figure 12 the distributions for the compact 
symmetric (cso) and lobe-dominated (ld) objects at 14.5 and 4.8 GHz. These groups
are comprised almost exclusively of galaxies, and, with the exception of 0723+679, 
they have steep total flux density spectra. For comparison, the distributions
are shown for the flat spectrum QSOs and BL Lacs in
Figure 13. The number of BL Lacs is small in the sample, but a distribution
based on a larger flux-limited sample of BL Lacs can be found in \citet{all99}.
We note that the QSO 0723+679 is included in 
both sets of distributions since its properties place it in both categories:
it is classed as an ld object based on its radio morphology while we find that the
spectral index is flat in the centimeter band. The csos are well-known to have
very low degrees of polarization at 4.8 GHz \citep{stan98}, and this result,
combined with recent VLBA observations identifying low values of the polarization 
at 8.4 GHz, have been used to argue that these objects are depolarized by Faraday 
rotation in a circumnuclear torus \citep{pec00}. This interpretation is consistent
with the wavelength-dependent fractional polarizations that we find.

\placefigure{fig12a}
\placefigure{fig12b}

\placefigure{fig13}

In Table 3 we summarize our results on polarization for the lobe-dominated
objects in this sample. The detailed polarization structure and spectral
variations within the lobes and central regions have been studied
with observations from the VLA and sophisticated analysis tools
\citep{kat99}. In the case of the FR Is, such data have revealed
sheath-like polarization structure in some objects \citep{har96}
and complicated spectral gradients in FR IIs \citep{tre01}.
However, our integrated measurements can be used to identify class
differences. We list optical class in column 2, and average fractional
polarization corrected for bias at 4.8 and 14.5 GHz in columns 3 and 4
respectively. An indicator of the spectrum of the polarized
flux (note that this is not fractional polarization) is given in column 5: these
are denoted by flat, inverted, or steep (F, I, S), based on visual inspection
of the longterm light curves. In some cases the spectrum has changed with time; 
in these cases (e.g. 0605+480) we list two classes, with the first denoting
the most characteristic spectral behavior during the full time window of the
study.  FR class is listed in column 6. For the FR II class objects we list
in column 7 the projected linear size taken from the compilation of  \citet{nils98}
which assumes $H$=50 km sec$^{-1}$Mpc$^{-1}$ and $q_o$=0.5. For the
FR Is we give projected length of the brighter jet originally tabulated in
\citet{bri84} but modified to a value of $H$=50 km sec$^{-1}$Mpc$^{-1}$ by \citet{har96}
and assuming $q_o$=0.5. 

\placetable{tbl-3}

In the three FR Is in the PR sample the spectra of the
polarized flux are steep.
In the case of the FR IIs we find a range of
behaviors: half of the sources have flat-to-inverted
polarization spectra based on the integrated measurements, while the remaining
half have steep polarization spectra. We have tried without success to
identify a common property from our own and published
data, for example source length, which could explain this dichotomy in
spectral behavior among a single class. Detailed 
rotation measure maps and results using spectral tomography have 
only recently become available for some member
sources. In the case of 3C 295 (1409+524), a source characteristically
exhibiting an inverted polarization spectrum, a VLA
study \citep{per91} plausibly attributed the
observed behavior to an intervening screen: Faraday depolarization
produced in thermal gas surrounding the source. However, it is not
clear whether this explanation can explain the range of behavior in other
sources in our study and the temporal variations we find.

\subsection{Preferred EVPA }

Inspection of the information on EVPA given in Table 2 shows that
longterm stability ($\ge 75 \%$ of the time) at 4.8 GHz was found in 28 
of the 62 sources, of which 12 are galaxies. In contrast, at 14.5~GHz only
7 sources show a statistically 
significant preferred EVPA $\ge 75 \%$ of the time;
however, several sources show preferred EVPAs for more than a third of
the time. Of these 7 objects with 
longterm, stable EVPAs, 5 are variable QSOs, one is a variable BL Lac and only one
is a steep-spectrum, quiescent, lobe-dominated galaxy. For the highly variable
sources, in particular the BL Lacs, we had attributed the lack of
long-term preferred EVPA at 14.5 GHz to the fact that the integrated
polarization is dominated by contributions from evolving source components
near, or in, the core \citep{all99}. However, we find here that even among the
BL~Lacs, long-term preferred EVPAs do exist over significant portions of the
time window, and over time periods which include several
events. This suggests that the magnetic field may have a `memory',
in the central radio-band regions, and that subsequent outbursts
in a source may exhibit the same or a related magnetic
field structure, or, alternatively, that there is
a persistent standing shock in the `core'.

Figure 14, showing the data for QSO 0212+738, illustrates another behavior we find 
in some AGNs: a preferred
EVPA which changes abruptly to a second longterm preferred EVPA.
Such behavior has previously been identified in non-sample members 1308+326 \citep{all99}
and 3C 279 \citep{all96}.
In 0212+738 we find a longterm preferred EVPA 
at 4.8 GHz  which is maintained in spite of the nearly continuous
activity in the source. The average centimeter-band spectral index
is very flat: $\alpha=0.03$. At 14.5 GHz, however, we find two
preferred orientations: a first near to the
value preferred at 4.8 GHz until 1995, followed by a rotation, and
a second relatively stable orientation nearly orthogonal to the
earlier value. 
This frequency-dependent EVPA separation has continued to the present time. We
attribute this behavior to a change in the opacity in the cm-band emitting region 
indicated by the evolution of the spectrum of the total flux density: the
source has become transparent, so that at 14.5 GHz we are looking further into
the jet at a region which has a substantially different magnetic field alignment. The
43 GHz map  of \citet{lis01a} at epoch 1999.26 shows four jet components, 
significant polarization
from both the core and jet, and a total polarization of 4.9\%. While
our data represent a sum of contributions from all components and extended
structure, the change we see is consistent with increased emission from
the strongest VLBI component. Both the millimeter
VLBA data and single dish centimeter-band data suggest that at this epoch the 
predominant magnetic field is aligned nearly perpendicular 
to the flow  direction of $\sim 111^{\circ}$. The behavior illustrates that significant
 changes in the characteristic behavior of the polarization, as well as
of the flux, can occur with time scales of a decade and that shorter
time windows (even a decade long) may not adequately sample the range of variability. 
In the case of this source, a sequence of well-sampled high frequency VLBP maps
would have been helpful in unraveling the complicated behavior we find.

\placefigure{fig14}

 \subsection{Relation of Magnetic Field Orientation to Flow Direction: Class-Dependent
 Differences}  

In Paper I we compared the preferred intrinsic EVPAs at 14.5 and 4.8 GHz 
with the flow direction ($\theta$) in the jet indicated by the morphology apparent in the
5~GHz Pearson-Readhead maps \citep{per88}. We found that at 14.5~GHz
there was no difference in distribution of $\vert$EVPA-$\theta \vert$
between QSOs and BL Lacs, but at 4.8 GHz the values of $\vert$EVPA-$\theta\vert$
clustered near  0-20$^\circ$ for the BL Lacs and were preferentially near 90$^\circ$ 
for the QSOs; the disparity between the shape of the distribution of orientation
difference for each optical class based on
our data at 4.8 GHz suggested intrinsic differences in the physical properties of the
objects such that the magnetic field in the emitting region is 
preferentially along the flow direction in QSOs and perpendicular to it in BL Lacs. 
This result supports the class differences in EVPA orientation identified in \citet{caw93},
but differs from it in concept since that result is based on single-epoch observations
of jet (and core) components with typical lifetimes of order a decade \citep{gab94}
while our result is based on longterm data, which may include many different
evolving components (which might have exhibited quite different EVPAs
from event to event), and with the major 
contribution to the integrated polarization
most likely arising in the core and innermost jet components at 14.5 GHz \citep{hom02}. 

With the substantial data set and higher resolution VLBI maps
now available, we reexamine the question of the
class-dependence of the magnetic field orientation. Inspection of Table 2 shows 
that there is a range of values for the flow direction found from the maps at the
various wavelengths for each source, which can be as large as tens of degrees. This
may result from curvature in the jet flow combined with scale differences in the regions
probed with the different observing frequencies, complicating the selection
of the most appropriate value for a comparison between the flow direction and
the EVPA. Further, the preferred polarization 
orientations found at 14.5 and at 4.8 GHz differ by several degrees in
some sources. That difference may result from opacity effects within the source 
so that the integrated emission arises from slightly different regions, with
differing magnetic field orientations. Alternatively the frequency-dependent
spread in preferred 
EVPA may indicate that our adopted rotation measure is incorrect in some sources.
Rotation measure mapping of several objects, \citep{tay00a,mut00,rey01},
has demonstrated that the rotation measure for the core can be very high
and that it can vary significantly from epoch to epoch. We have
assumed that the  rotation measure is spatially and temporally invariant. All
of the above factors combine to complicate a comparison
of the magnetic field orientations, deduced from the polarization, 
with the flow direction in the jet.

In Table 4 we show a grid of values comparing EVPA and $\theta_{\nu}$
for the BL Lacs of the sample. Note that there is an ambiguity of 180$^{\circ}$
in the determination of the EVPAs and that we have
restricted the orientation difference to lie within a 90$^\circ$ range. 
We find that the orientation difference, $\vert$EVPA-$\theta\vert$, 
is generally $\leq 36^{\circ}$  (see discussion of 3C 371 below) confirming
the result from Paper I that in BL Lacs the magnetic field orientation (orthogonal
to the EVPA) is preferentially aligned transversely to the flow direction. 
\citet{lis01a} has looked at the distribution of the 
EVPA of the core with respect to the innermost jet direction in BL Lacs using
single epoch 43 GHz VLBI observations and also finds that these offsets are confined
to the range $\leq 40^{\circ}$. That these distributions are similar is consistent with
Lister's conclusion that the bulk of the polarized emission on parsec scales in flat
spectrum members of the PR sample comes from the core component. This result appears
to imply that there is a preferred EVPA of the core component over the longterm, but that 
this can change with time as unresolved components emerge with different orientations.
Earlier results at 5 GHz by Gabuzda and coworkers, e.g., \citet{gab92}, also showed
systematic differences in the polarization structure in BL Lacs and QSOs, but these
were based on the distributions for jet components, and not for the cores, which 
showed no clear preferred range of orientation.

\placetable{tbl-4}

In Figure 15a we compare the orientation difference between the preferred 
intrinsic EVPA at 4.8 GHz ($\chi_0$ in Table 3) and the flow direction at the
highest VLB frequency (generally 43 GHz,
but supplemented by observations at 15~GHz for three galaxies) for all 
PR sample sources with both a well-defined flow direction
and a preferred EVPA. The objects have been separated by optical class. 
We chose this frequency because of the larger number of sources which
exhibit preferred EVPAs, and for comparison with earlier work based
on VLBP observations at this frequency.
The distribution
is consistent with a systematic difference between the behavior for QSOs and
for BL Lacs such that the dominant persistent magnetic field orientation is
along the flow direction in QSOs (orientation difference near $90^{\circ}$)
and perpendicular to it in BL Lacs. We note that there is a large range of
values for the QSOs; for the BL Lacs the values are closely clustered. An exception
is the BL~Lac object
3C~371 which shows a large orientation difference characteristic of a
QSO; however, in this source Lister has found no polarization in the core at
43 GHz, in agreement with the earlier 5 GHz result of \citet{gab89},
suggesting that our measured polarization originates in extended structure. 
The suggestion is consistent with the arcsecond scale polarization mapping
of the source at 1.4 GHz by \citet{stan97}. A KS test of
the distributions for the BL Lacs and QSOs gives a probability of 3.97$\%$
that they are drawn from the same parent population. 
A $\chi^2$ test of the distribution for the QSOs gives a probability of 2.1$\%$
that this distribution is random. Thus, our result is suggestive at the $2\sigma$
level of two different 
populations. Within the shock-in-jet scenario this 
would be consistent with differences in the allowed range of obliquities relative
to the flow direction, or alternatively to the increased importance of a longitudinal
magnetic field component in QSO flows, possibly related to
shear. These alternatives are discussed in section 5.

\placefigure{fig15}

Because the polarization measurements at 4.8 GHz may contain contributions from
unresolved components near the core, we show in Figure 15b the same plot based
on the preferred intrinsic EVPAs
at 14.5 GHz. We have included all sources for which a significant
preferred EVPA was identified; in general these are maintained for
shorter time intervals than at 4.8 GHz. All of the BL Lacs in the sample exhibit
preferred EVPAs at 14.5 GHz, as well as at 4.8 GHz. However, nine QSOs which exhibited
preferred EVPAs at 4.8 GHz do not exhibit them at 14.5 GHz, with the converse for one
source: the ssc object 0538+498 which has very weak polarization at 4.8 GHz. Unfortunately,
there are only 21 sources with both preferred EVPAs and well-identified flow directions at
14.5 GHz which weakens statistical tests due to small numbers. Nevertheless, while the
distribution for the QSOs now more closely resembles a random one, the BL Lacs remain
clustered near small orientation differences, with the exception of 3C~371 already discussed.

 \subsection{Systematic Temporal Changes in EVPA}
We have examined the temporal variations in EVPA to look for 
behavior which might be expected as
a signature of a precessing jet, and were unable to identify such behavior.
Prime candidates might be the compact symmetric objects, 
which are believed to be young and possibly precessing based on their unusual
VLBI morphologies \citep{tay96}; however, as already discussed, the polarization
is relatively low in these objects.  Furthermore, if geometric precession were to occur 
in the jets of some sample members,
the expected systematic variations in EVPA might be masked by the changes in 
the evolving source components contributing to our integrated
polarization. As an added complication, the expected precession period even
in the relatively rapid variable, OJ 287, is estimated to be 130 years
\citep{leh96} and even longer in other objects \citep{kat97},
while our observational time window is only of order a few decades. Thus,
it is uncertain whether the expected signature could, in practice, be
identified. 

 \subsection{Results of Search for Periodicity}

 We have carried out Scargle periodogram analyses 
on all sources in the sample and find no
{\it strong} evidence for periodicity in any of the sample sources based
on our data. Evidence for quasiperiodicity in the radioband 
light curves for sample member DA 55, based on a time series analysis, has been carried
out by \citet{pya02}, and both the radio structure and broadband spectral evolution suggest
helical jet structure in a second sample source 0954+658 \citep{rai99}. Past attempts
 to identify periodicity in the centimeter band 
emission from UMRAO data alone, in even the most active AGNs, using both Scargle periodograms
and Morlet wavelets, have been hampered by the fact that large, discrete events
occur typically only once every few years. Hence, while there have been
hints of periodicity in the analysis for a few objects exhibiting several events, e.g.
\citet{roy00}, these  determinations have not had high significance 
except in the case of OJ~287 \citep{hug98} which also remains the
best case  for precession \citep{val00}. A longer timebase increases the
prospects of identifying periodic behavior, if present, in any of these sources.

The Scargle periodogram analysis identifies marginally
significant power peaks in four sources (0804+499, 0954+658, 1637+574, \& 1739+522),
but these may be tied to the window function or be
cases of aliasing, and a longer time period of observations will be
required for verification. For two of these four objects an independent structure function
analysis yielded unusually short characteristic
timescales (0.63 years for 1739+522 and 1.1 year for 1637+574) compared to the
average value of $\sim$ 2 years found from analysis of a sample of highly
active objects \citep{hug92} suggesting that further observation of these
selected objects is warranted and that they may be special cases
exhibiting unusual variability properties.
For 0804+499  a structure function analysis identifies 
wavelength-dependent characteristic timescales and the shortest time scale
we have found for any source: 0.15 years at 14.5 GHz. This source is an 
intraday variable \citep{wag91}, but our data probe a different time domain, 
identifying a time scale of order 2 months. The periodogram analysis also identifies
an unusually broad distribution of power in two sources (0016+731 and Mkn 501) relative
to the results for other sources and is consistent with white noise structure functions, perhaps
resulting from temporally undersampled, very short timescale activity. We
conclude that while the search for periodicity was inconclusive, it has identified unique, 
interesting objects worthy of further investigation. A recently completed cross wavelet
analysis of the data is presented in \citet{kelly02}. 

\section{DISCUSSION AND CONCLUSIONS}

 While our light curves have allowed us to delineate the
characteristics of the variations in the emission, understanding the evolution of
the jet flow, its origin, and its relationship to the putative black hole which
fuels it can best be accomplished by a combination of well-sampled light curves
and VLBI/P images at several frequencies, with adequate time resolution to follow 
the evolution of individual features. Such maps are now  available for limited
time periods for a handful of sources, selectively chosen to be blazar-type AGNs, e.g., 
\citet{hom01,hom02}, but, as demonstrated here, such AGNs represent only one
facet of the variability phenomenon. 

Within the framework of a shocked-jet
model, associating the variability we find in both total and polarized flux
with the development of shocks in the flow,
our results on magnetic field orientation suggest that in BL Lacs shocks develop
with a limited range of obliquity to the flow direction and are preferentially nearly 
transverse to the flow direction; in QSOs the range is larger and preferentially at
large angles to it suggesting that oblique shocks, rather than transverse ones,
form in these flows. It has been proposed that the longitudinal
magnetic field alignment observed in QSOs is due to shear \citep{caw93}.
However, while shear could produce an increase in polarization associated with the ordering
of the field, it would result in a very specific field orientation. Thus we prefer
an explanation invoking shocks with a range of obliquities to the flow direction, an
explanation which is consistent with the structural evolution seen in
time series of VLBA  polarization maps.
While shear, to some degree, is expected in the flows, either as a vortex sheet at the
jet-ambient medium interface, or in a shear layer permeating the emitting region, no
quantitative shear models for parsec-scale jets currently exist for specific
tests against the data. Additionally, we do not believe that the 
observational evidence cited in
favor of shear is unambiguous: 1) while VLBP maps of the QSO 1055+018 and of the
 BL Lac-type object 0820+225 show longitudinal sheath-like magnetic fields in
the parsec-scale flows, we do not believe that
there is conclusive evidence that these are associated with shear, nor do
these two cases explain the QSO-BL Lac dichotomy since one of the objects is a
BL Lac; 2) recent analysis of VLBP maps for 12
blazars \citep{hom02} observed during a one-year time period showed 
that the EVPAs of the jet features rotated with time such that the magnetic
field has become more aligned with the jet axis with increasing distance from the core,
but the origin of the change in net field orientation could not be unambiguously
identified: it is neither clear
whether an increased importance of shear with core distance
could produce such changes within the linear dimensions probed, nor was the
expected associated increased ordering of the field identified in
4 of the 5 cases. Temporal changes in EVPA are currently being studied within
the framework of
evolving oblique shock models using VLBA data obtained over a 30-month time
span, allowing a larger range of core distances to be explored, and 
including a wider range of frequencies \citep{all02}.

Precession has been discussed widely in the literature, and might be expected 
based on the range of morphologies seen in maps from VLB surveys \citep{kel98},
but we are unable to identify its expected signature in our data.
A combination of longterm monitoring data and VLBI mapping may
lead to more conclusive evidence, and the data included in a recent study of
the BL Lac object ON~231 
\citep{mas01} is an observational initial step in this direction. However, while
evidence supporting the helical character of jet flows is strong, 
this may result from instabilities in the flow rather than from precession \citep{har01}. 
Thus, in our view, a convincing case for precession has only been made for
a single source: OJ~287.

The millimeter imaging data now available, combined with sophisticated
analysis tools, permit studies of the evolution of the 
structure in the inner regions of the jet. With the aid of such modeling, it is now 
possible to follow the structural development of individual regions in 
both polarization and total flux, and to identify the complex changes in 
the flow direction, including bends and changes in the underlying magnetic field
direction on sub-milliarcsecond 
scales. These data probe quite different regions from those initially studied at 5 GHz, 
and both opacity effects and Faraday effects are reduced. Detailed model fitting in the
range 15 to 43 GHz has already demonstrated that the so-called `core', may, in 
fact, contain contributions from  newly-emerging, blended components, further
complicating an earlier simpler picture. Indeed, recent model fitting of the VLBP data 
for some core-dominated blazars, of both the BL Lac and QSO classes, demonstrate
quantitatively that
the major contribution to the integrated polarized flux comes from the `core' itself or
the next innermost component \citep{gab00,hom02}. Clearly, flux and
polarization monitoring
at well-selected frequencies remains an essential tool for probing the detailed 
changes in the jet flow and unraveling the complex 
opacity-dependent effects within it. Hopefully, a combination of imaging and 
monitoring of both blazars and other types of extragalactic objects will ultimately
lead to a more complete understanding of their origin and evolution.

Finally, while we continue to believe that the features of the variability
apparent in our data can most readily be explained within the standard picture 
of relativistic jet flows containing passive magnetic fields
and the naturally-developing instabilities within them, we note
that an alternative picture has recently been proposed by \citet{blan03} based on
dynamo models and ensuing currents. Once detailed predictions
of the flux and polarization are available for this class of model, it will be most
interesting to compare them with the range of behaviors discussed here.

Our main results are as follows:

 1. We identify variability in steep spectrum objects (ssc and cd classes by radio
morphology) which is characterized by infrequent events or longterm monotonic changes.
This group includes 3C~147 and calls into question its use as a secondary flux standard.

2. We have tentatively identified small-amplitude variations in several lobe-dominated
sources including 3C~179, 3C~236, and 3C~390.3. These variations
are consistent with information on core strengths and the ratio of 
core to total flux known from radio maps of those objects.

 3. We have provided evidence in support of our view that variability is a
pervasive phenomenon in extragalactic objects, and that the majority of extragalactic objects 
exhibit activity which can be identified from integrated light curves over
time periods of order 1-2 decades. This long time range is crucial for the
case of the steep spectrum objects where large events are required for detection
because of the dominant contribution from extended structure.

4. We find no strong evidence for periodicity for any sample members based on a
Scargle periodogram analysis of our total flux density data, nor are we able to 
identify the signature of precession in the temporal evolution of the EVPAs.

5. We find a range of behavior in the polarization spectra of the steep-spectrum
 objects. The flat-to-inverted spectra we find in several source members is consistent with
  Faraday depolarization by a Faraday screen. 

6. We find longterm EVPA stability in many objects indicative of a persistent, dominant
magnetic field orientation. When compared with flow directions from VLBI
morphology, the field directions are consistent with class-dependent differences between
BL Lac and QSO flows. 


\acknowledgments
This work has been supported in part by NSF grants AST-8815678, AST-9120224,
AST-9421979, and AST-9900723.  We thank A. Polatidis 
for providing unpublished jet orientations at 6 cm, H. Falcke for useful
suggestions regarding the identification of variability in lobe-dominated sources,
M. Lister for providing results in advance of publication and for insightful
comments, and an anonymous referee for comments which improved the content and
presentation of the paper. We gratefully thank J. A. 
Zensus both for permission to quote results based on unpublished maps from the 2 
cm VLBA survey (MOJAVE),
and for his hospitality at MPIfR to M.F.A. and H.D.A. where part of this research was 
carried out. Finally, M.F.A. and H.D.A. thank NRAO for its hospitality during which 
this work was completed.
This research made use of the NASA/IPAC Extragalactic Database (NED) which is operated
by the Jet Propulsion Laboratory, California Institute of Technology under
contract with NASA. The operation of the 26-meter
telescope is supported by the University of Michigan Department of Astronomy.

%
%

\clearpage
\begin{figure}
\plotone{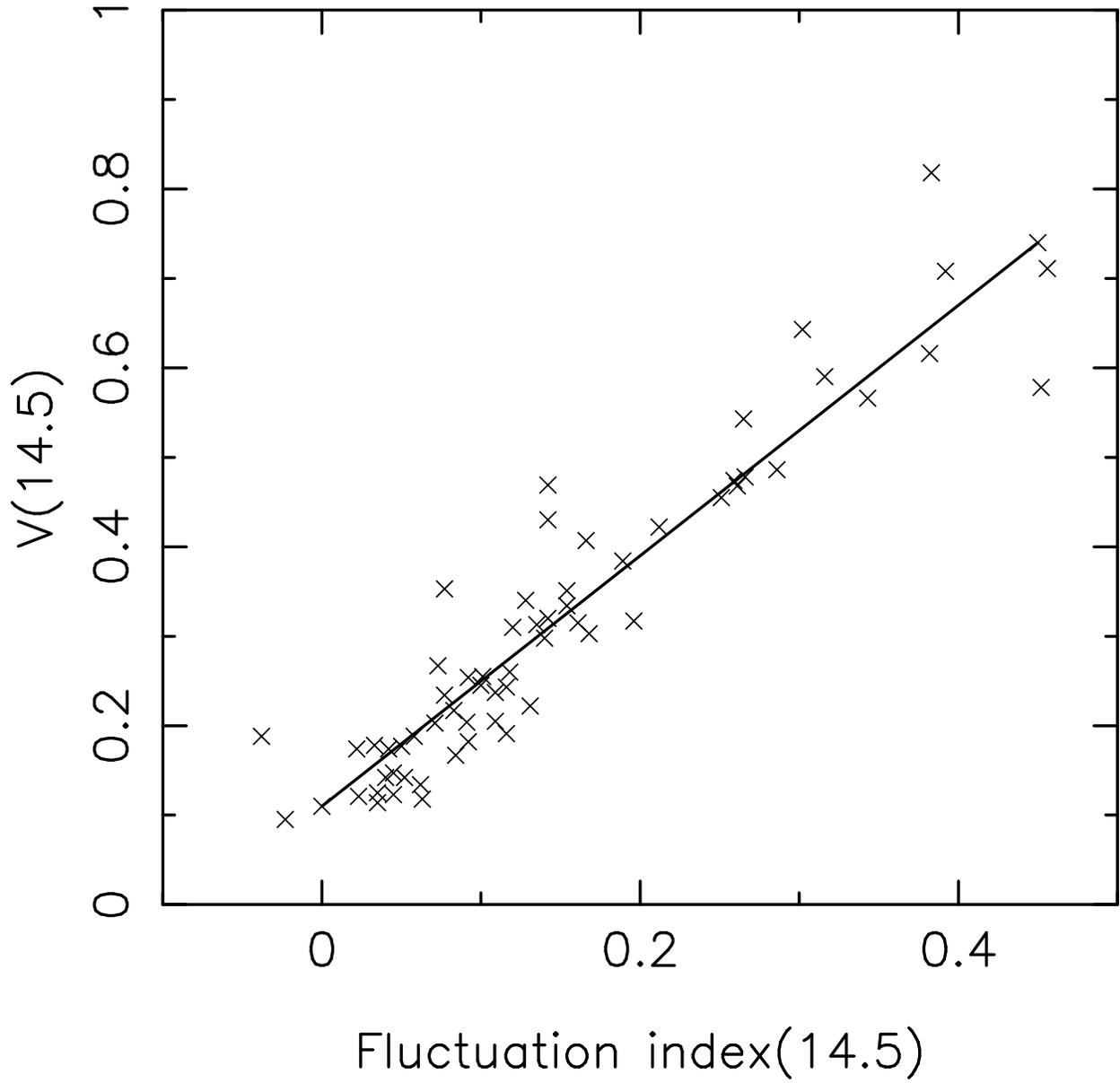}
\caption {Variability index versus fluctuation index based on data during
1984.58-2001.25. The line denotes a least squares fit to the data.
The correlation coefficient is 0.937. The source with the unusually high
amplitude variability is the BL~Lac object 0954+658.
\label{fig1}}
\end{figure}

\begin{figure}
\plotone{f2.eps}
\caption{Fluctuation index versus redshift. The objects have been coded by optical class. 
Note that with the exception of 3C~84, the galaxies have FI(14.5)$\leq0.16$. \label{fig2}}
\end{figure}

\begin{figure}
\plottwo{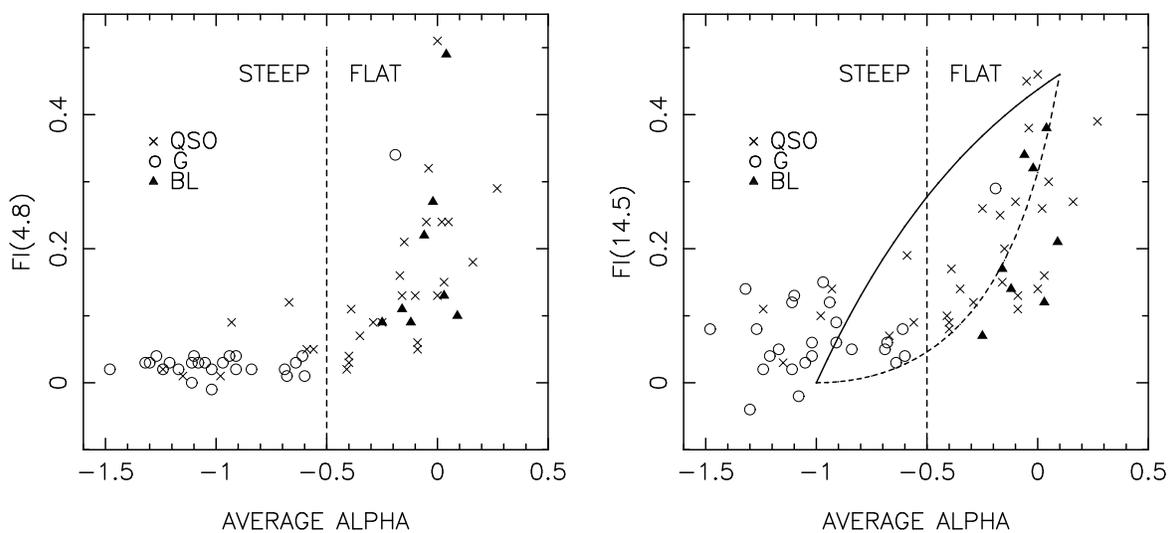}{f3b.eps}
\caption{Left: fluctuation index at 4.8 GHz versus average alpha (14.5-4.8 GHz) based on
data in the time interval of this paper: 1984.58-2001.25. The objects have been coded by
optical class, and the demarcation between flat and steep spectrum objects is
marked by the dashed vertical line. Right: fluctuation index at 14.5 GHz versus alpha.
The full and dashed curves shows the locii for two assumed two-component models 
as described in the text.\label{fig3}}
\end{figure}

\begin{figure}
\plotone{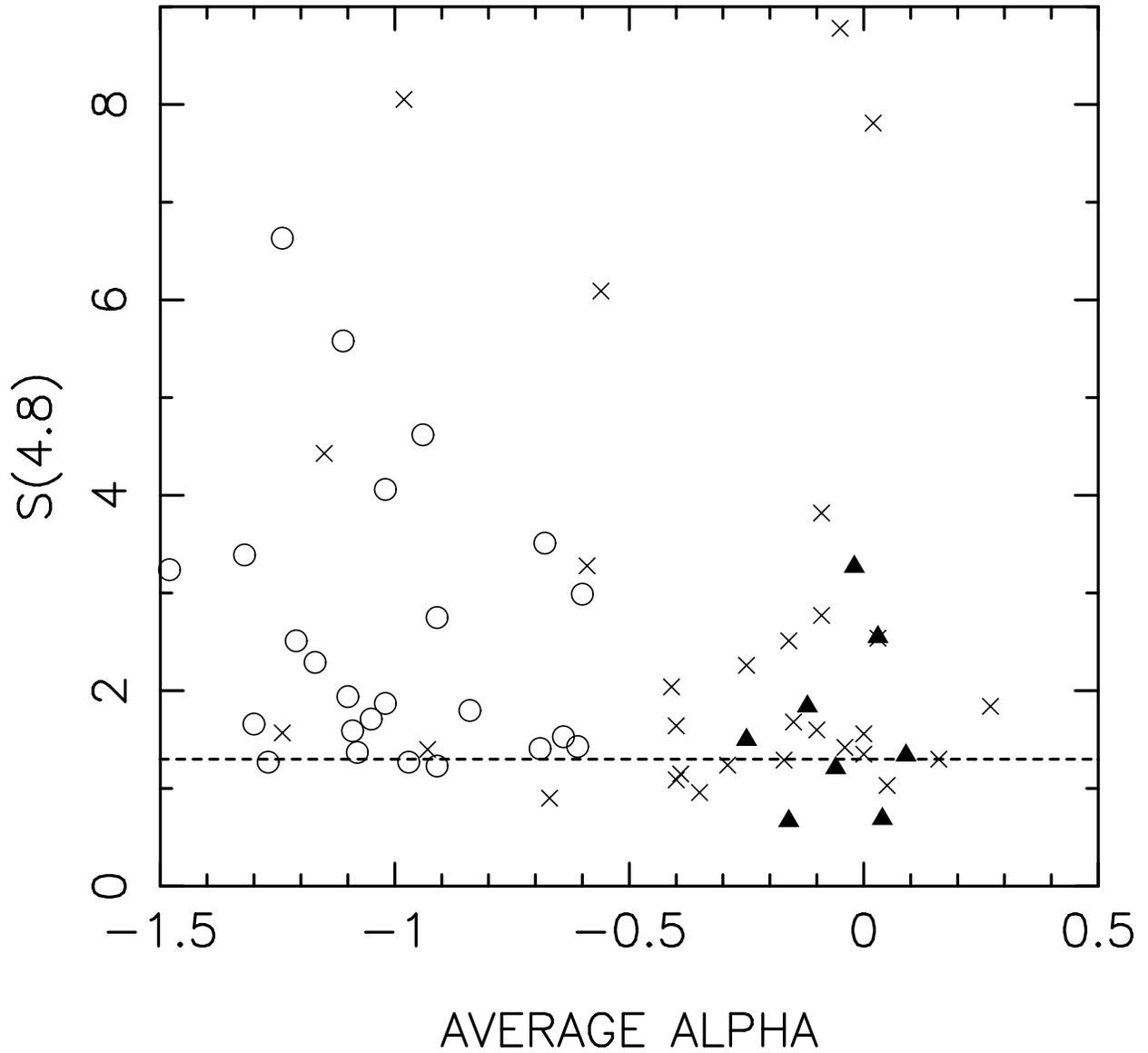}
\caption{Average 4.8 GHz flux versus average spectral index during 1984.58-2001.25.
The dotted horizontal line marks 1.3 Jy, the 5~GHz flux density used for source
selection. Sources below this line would not have been included in the survey based 
on their average flux density. The very bright source 3C~84 lies outside the flux
range shown.
\label{fig4}}
\end{figure}

\begin{figure}
\plotone{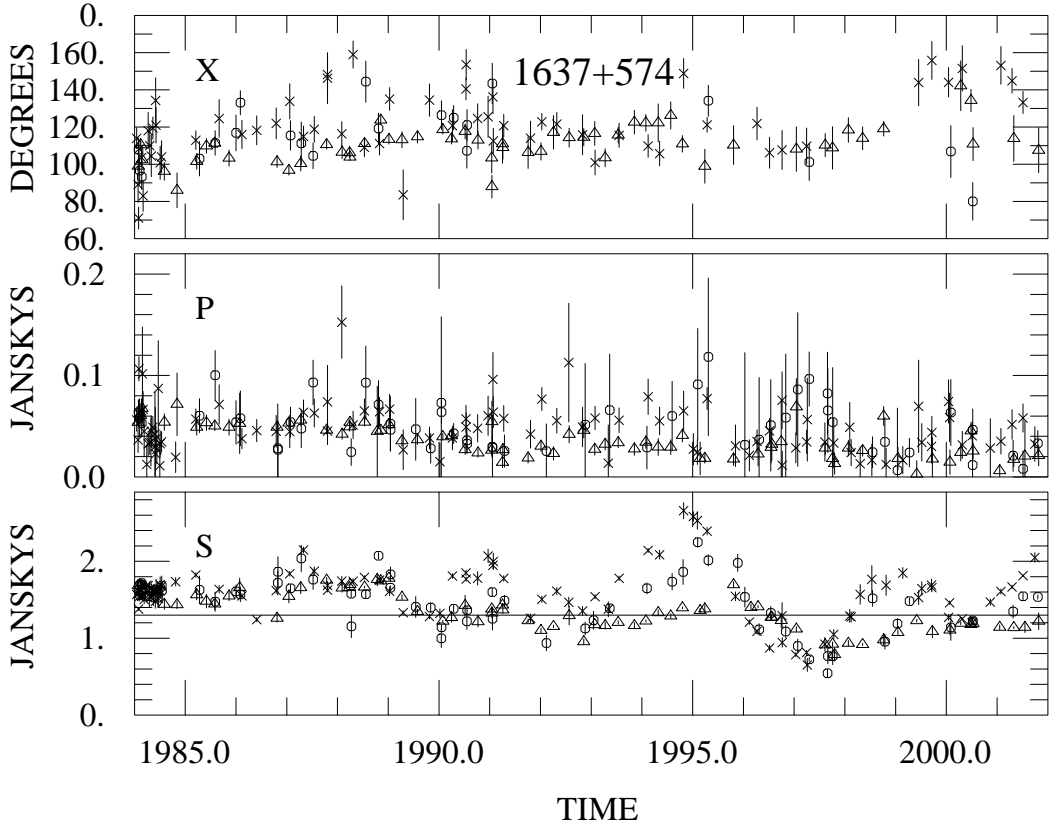}
\caption{From bottom to top, daily averages of the total flux density,
 polarized flux, and EVPA for the flat spectrum QSO 1637+574. The observations at 14.5, 8.0
and 4.8~GHz are  denoted by Xs, circles, and triangles respectively,
and the polarization has been corrected for Faraday rotation assuming
a rotation measure of +22 rad m$^{-2}$. The symbol
convention is adopted in all of the subsequent light curves.
The solid line in the lower panel 
marks the 1.3 Jy total flux density level used as a selection criterion
for the sample. The source has exhibited a series of well-defined events
occurring every few years with a flat to
inverted spectrum. \label{fig5}}
\end{figure}

\begin{figure}
\plotone{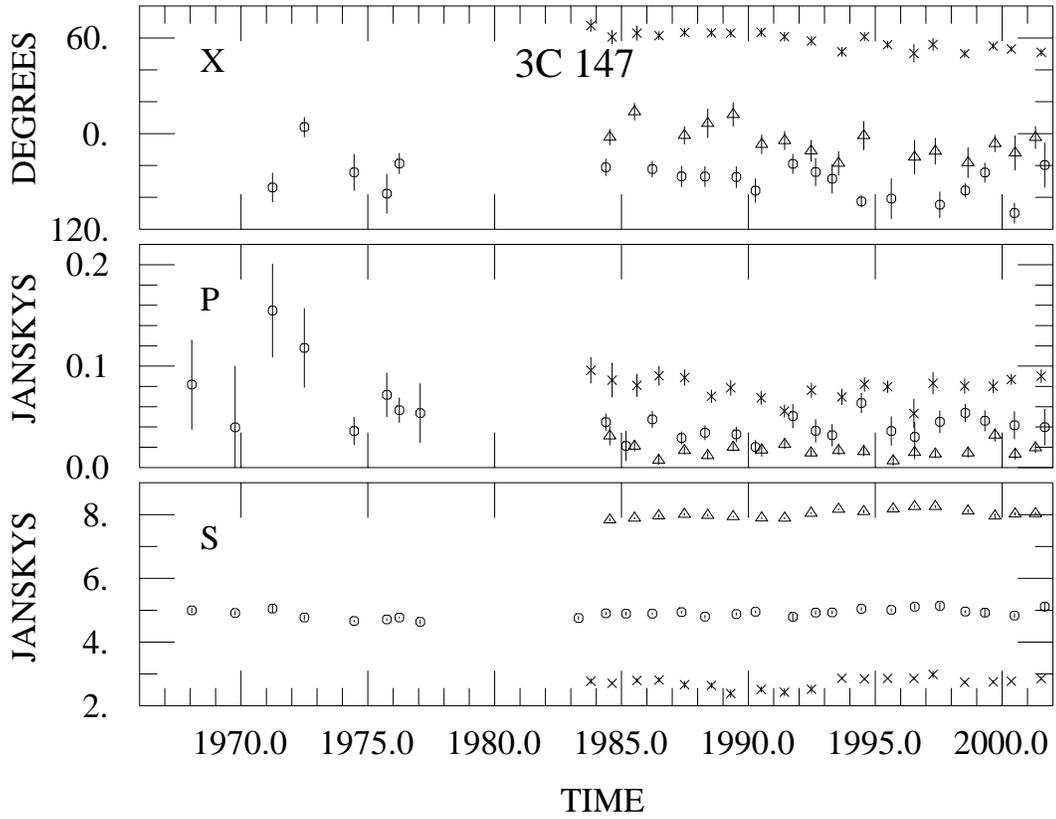}
\caption{Yearly averages of the total flux density,
 polarized flux, and EVPA for the steep spectrum QSO 0538+498 (3C 147).
 The symbols are as denoted in Figure 5. \label{fig6}}
\end{figure}

\begin{figure}
\plotone{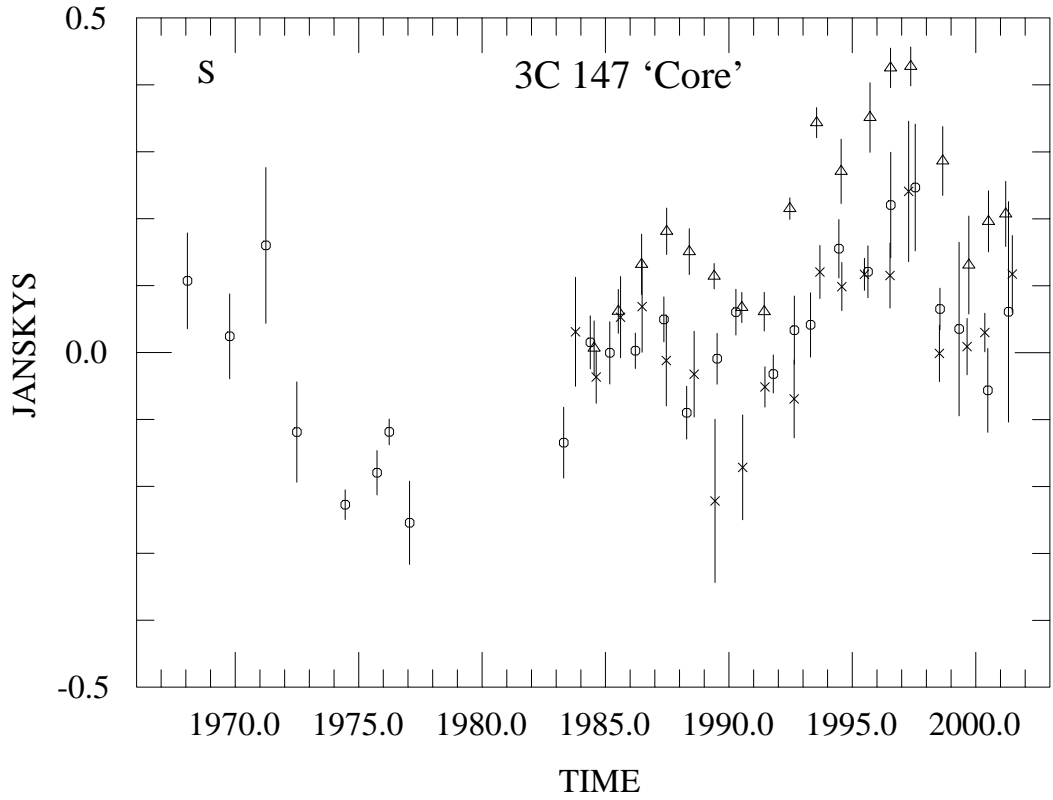} 
\caption{ A blow-up showing flux density only for 3C~147, with baseline fluxes removed
as specified by the spectrum given in \citet{ott94}.\label{fig7}}
\end{figure}

\begin{figure}
\plotone{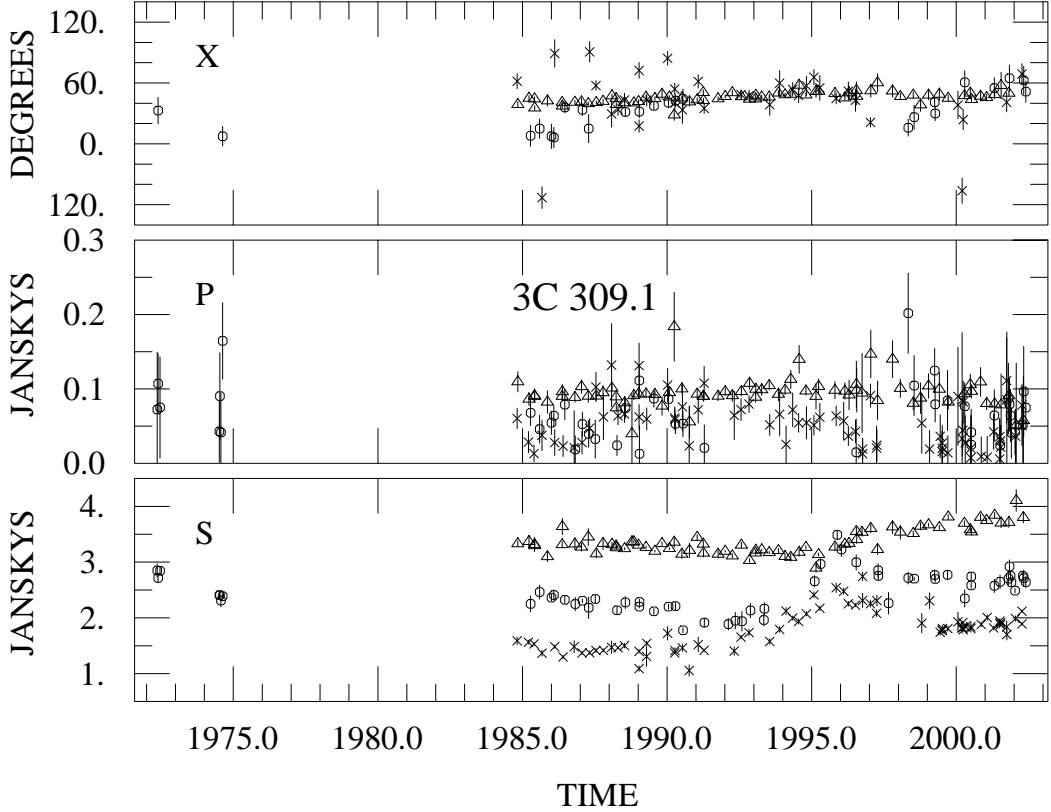}
\caption{From bottom to top, daily averages of the total flux density,
 polarized flux, and EVPA for the steep spectrum compact object
1458+718 (3C 309.1). The full time range of our data is shown.
Although undersampled, variations clearly occurred during the mid 1970s,
as well as in the mid 1990s. The polarization has been corrected
for Faraday rotation assuming a rotation measure of 75 rad m$^{-2}$.
This integrated RM differs slightly from the value of 60 rad m$^{-2}$
given by \citet{aar98} and discussed in the text.\label{fig8}}
\end{figure}

\begin{figure}
\plotone{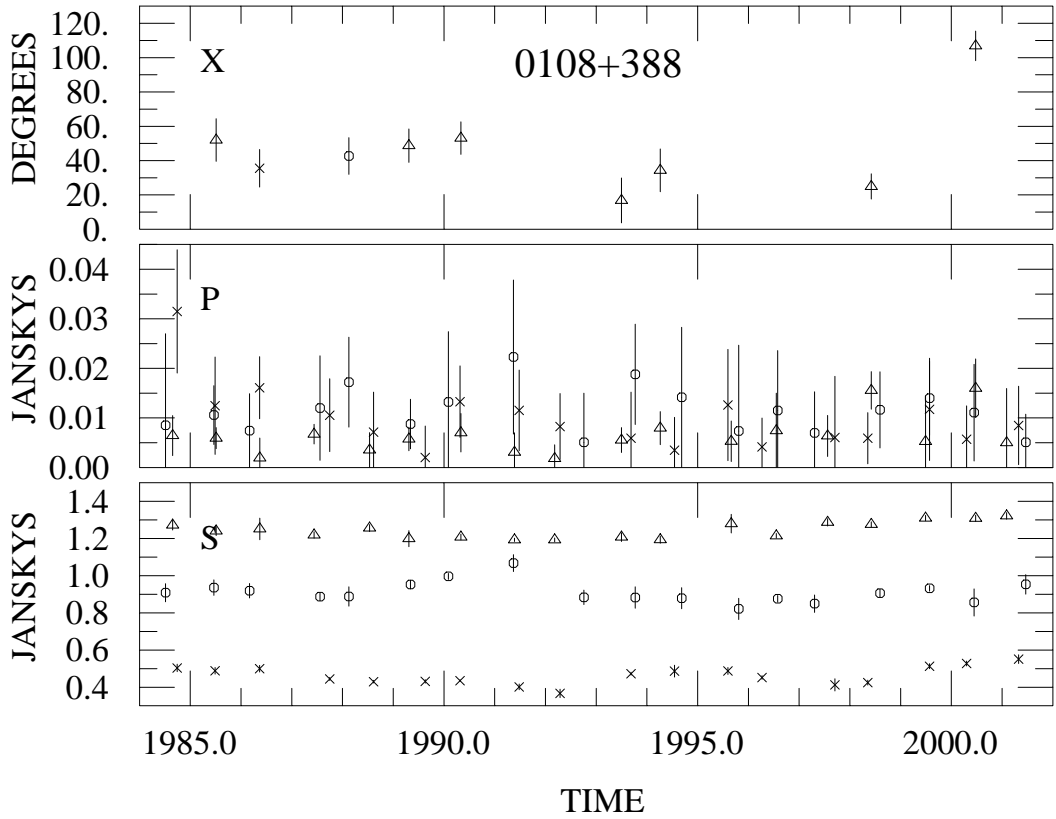}
\caption{From bottom to top, yearly averages of the total flux density,
polarized flux, and EVPA for the compact symmetric, GHz-peaked
galaxy 0108+388. The low degree of variability is typical of that seen
in other csos, and the total flux density spectrum is characteristic
of GHz-peaked objects. The spectral turnover at 3.9 GHz in this
object \citep{stan98}, is outside the range of our spectral
coverage. \label{fig9}}
\end{figure}

\begin{figure}
\plotone{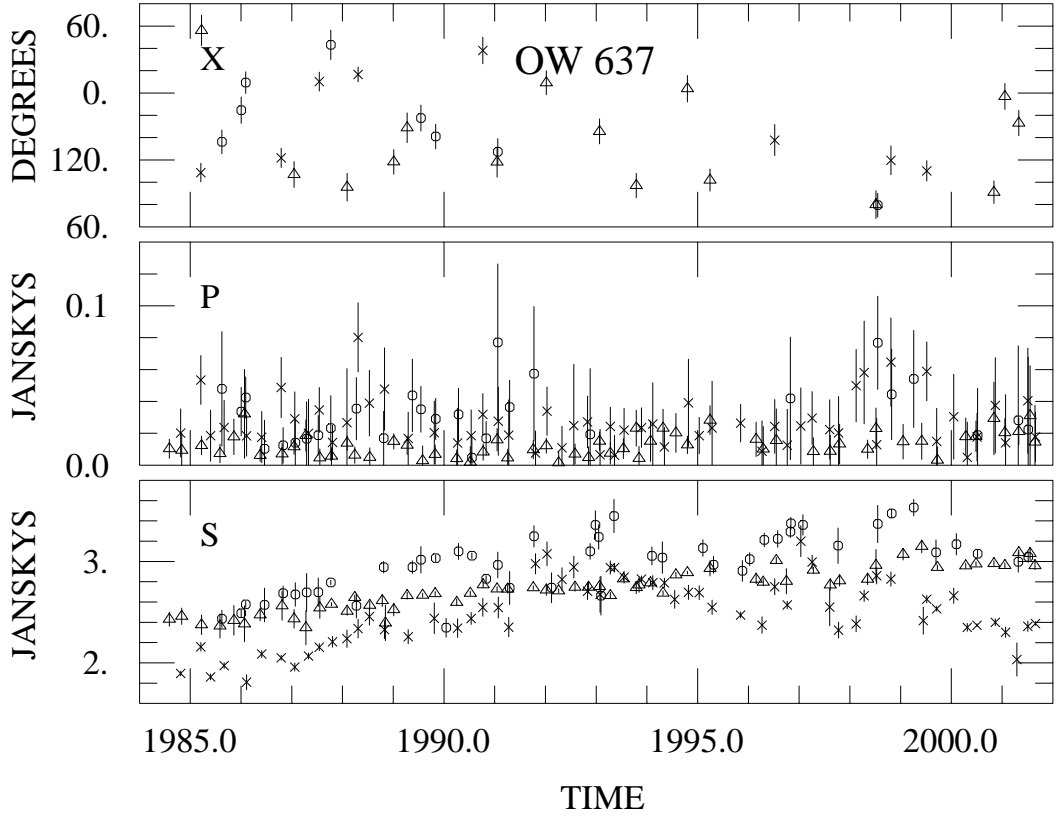}
\caption{From bottom to top, monthly averages of the total flux density,
polarized flux, and EVPA for the COINS sample member 2021+614. The polarization
has been corrected for Faraday rotation assuming a rotation measure
of -99 rad m$^{-2}$; few values of EVPA are shown because most are below our
signal-to-noise criterion for inclusion. The spectral turnover,
at 8.4 GHz \citep{stan98}, is within our spectral coverage. The source
exhibits well-defined variability while maintaining its GPS spectral shape.\label{fig10}}
\end{figure}

\begin{figure}
\plotone{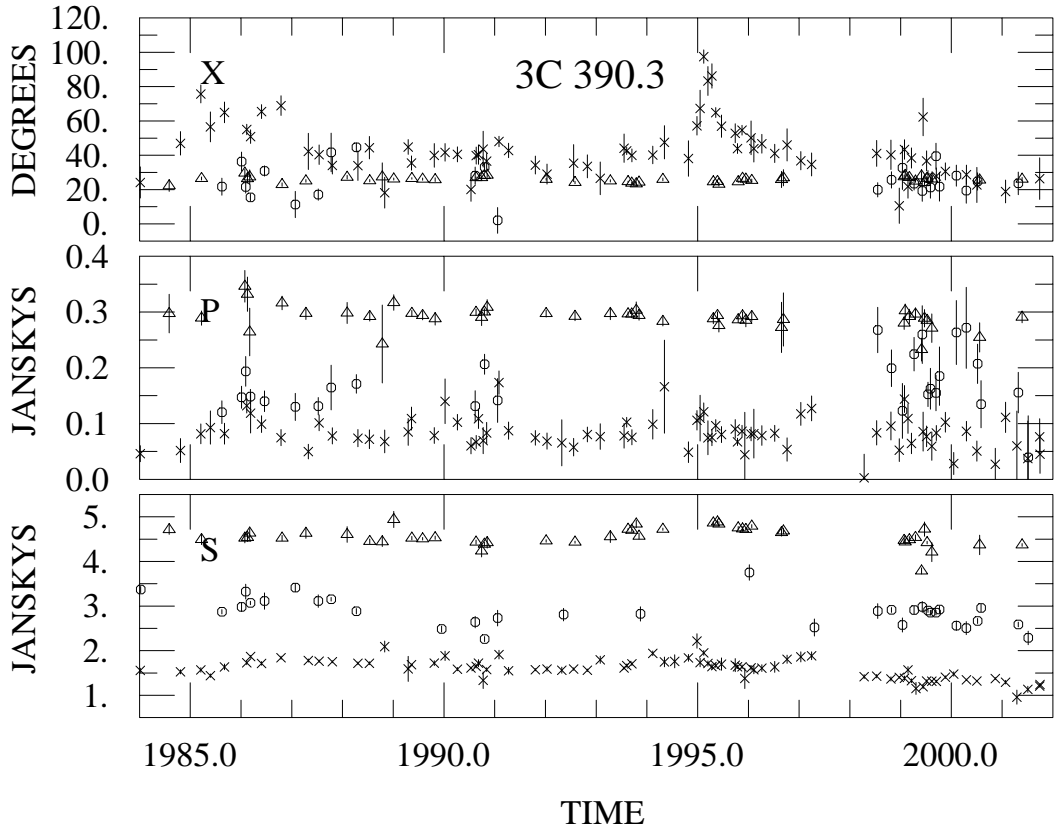}
\caption{From bottom to top, monthly averages of the total flux density,
polarized flux, and EVPA for the X-ray bright galaxy 3C~390.3.
The EVPA (top panel) has remained stable at 4.8 GHz, but it shows at 
least one large swing at 14.5~GHz. The monthly averaging
has smoothed out the short term activity at 14.5~GHz
in both total flux density and polarization. \label{fig11}}
\end{figure}

\begin{figure}
\plottwo{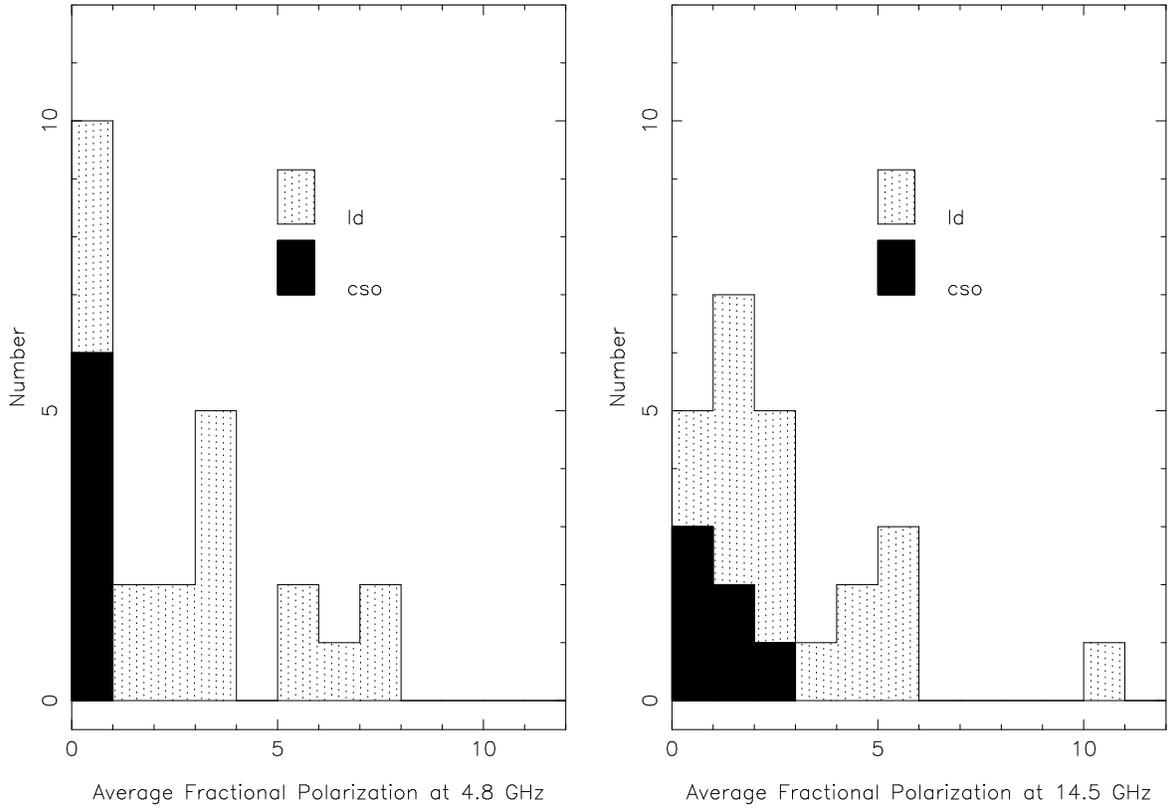}{f12b.eps}
\caption{Histogram of average fractional polarization based on the data at 4.8 GHz
(left) and at 14.5 GHz (right) for the steep spectrum radio classes compact
symmetric (cso) and lobe-dominated (ld). \label{fig12}}
\end{figure}

\begin{figure}
\plottwo{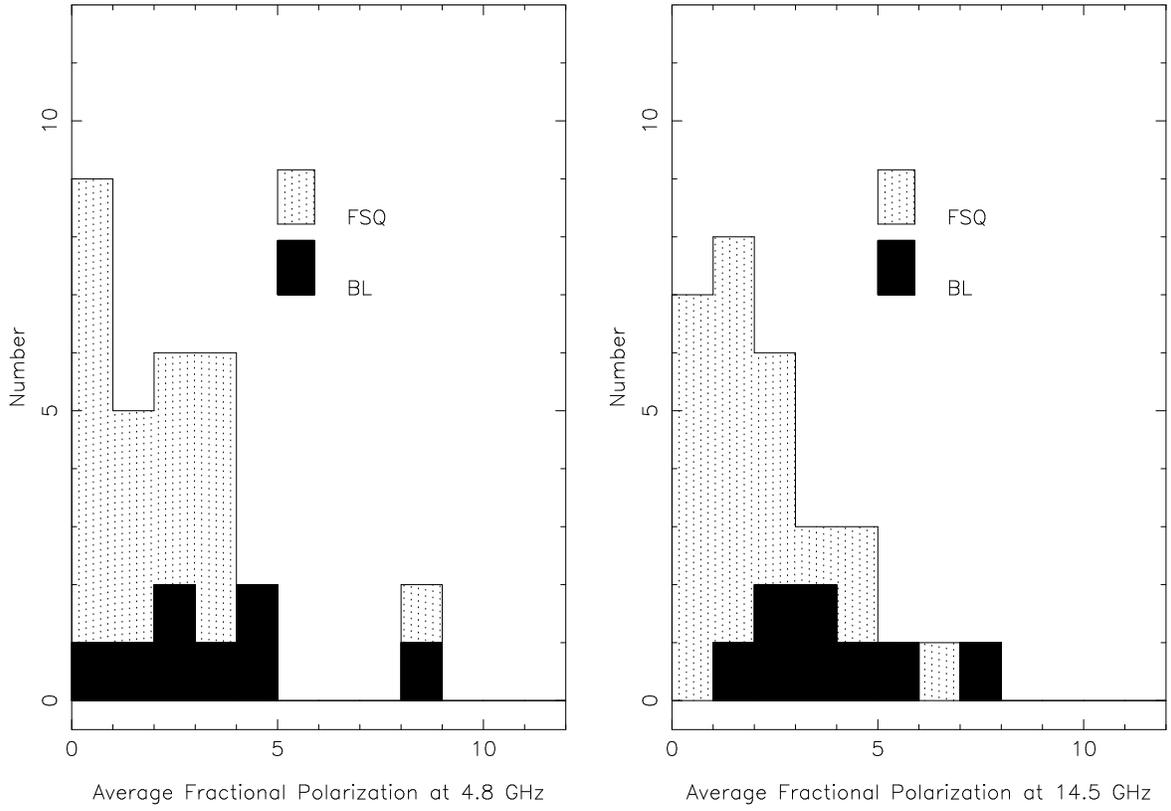}{f13b.eps}
\caption{Histogram of average fractional polarization based on the
data at 4.8 GHz (left) and 14.5 GHz (right) for the flat spectrum
QSOs and BL Lacs. \label{fig13}}
\end{figure}

\begin{figure}
\plotone{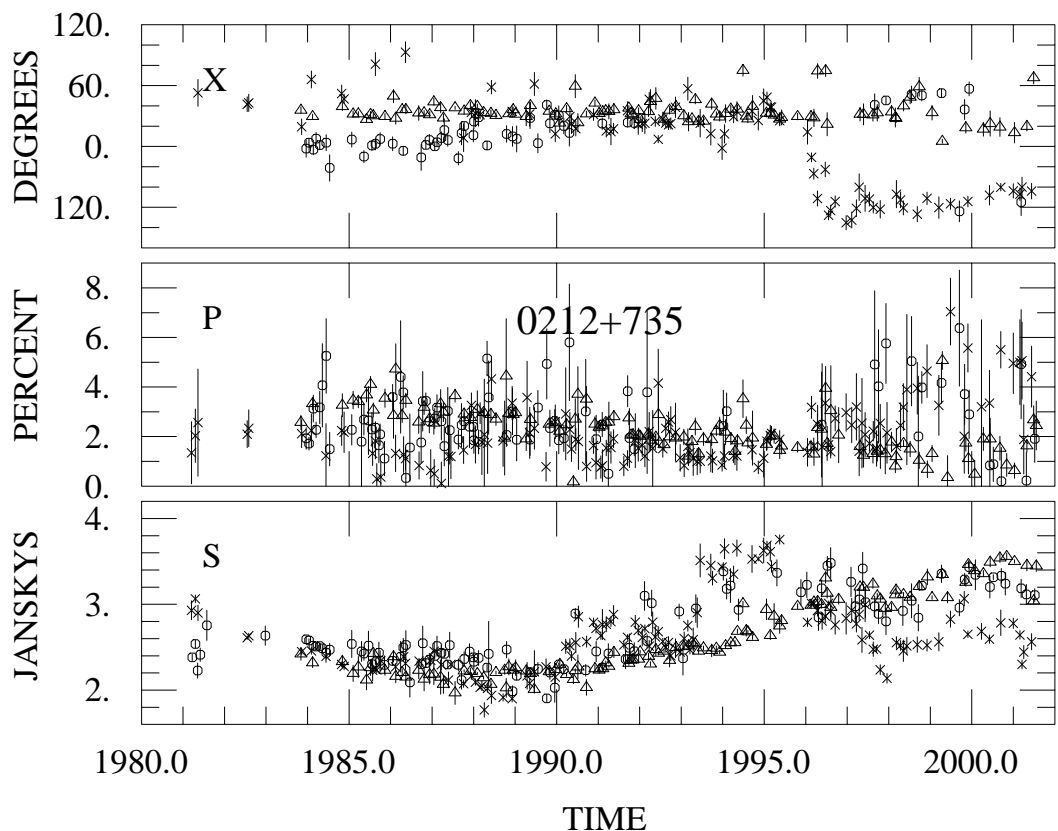}
\epsscale{0.6}
\caption{Monthly averages of the total flux, fractional polarization, and
EVPA for the  variable, high-redshift QSO 0212+738. A dramatic change in the
behavior of the EVPA commenced in 1996 when the orientation of the EVPA at
14.5~GHz aligned with the
orientation of the flow direction (B perpendicular to the flow). During this
time period the fractional polarization increased significantly at 
8.0 and 14.5 GHz and decreased at 4.8 GHz. Note also the decrease in the
total flux at 14.5 GHz during the same time period.
\label{fig14}}
\end{figure}

\begin{figure}
\plottwo{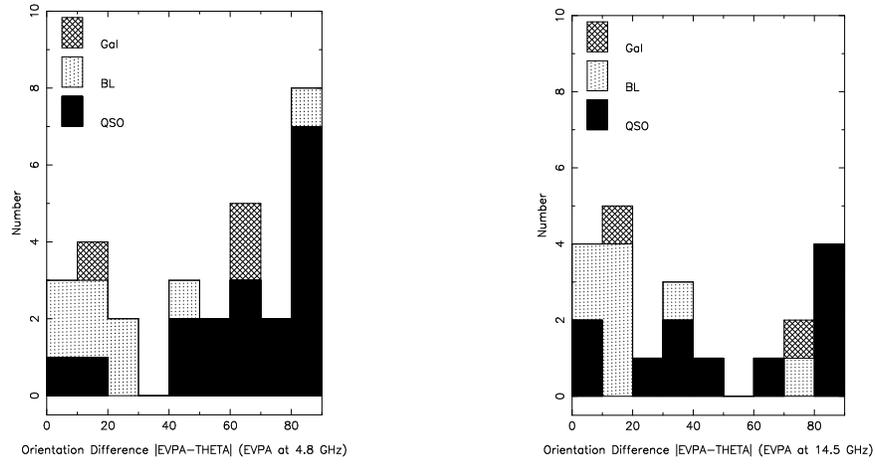}{f15b.eps}
\caption{Left: orientation difference between the preferred EVPA at 4.8 GHz and
the flow direction indicated by the observation of $\theta$ at the highest
VLBI frequency given in Table 2 for each source. Right: orientation difference based on the
preferred EVPA at 14.5 GHz and the same adopted flow direction. In a transparent synchrotron 
source the EVPA is orthogonal to the magnetic field direction. \label{fig15}}
\end{figure}

\clearpage

\begin{deluxetable}{clcclcrrrl}
\singlespace
\tabletypesize{\scriptsize}
\tablecaption{SELECTED EMISSION PROPERTIES OF PROGRAM SOURCES. \label{tbl-1}}
\tablewidth{0pt}
\tablenum{1}
\tablehead{\colhead{Source} & \colhead{Name} &  \colhead{Opt.} &  \colhead{radio}  & \colhead{z}
 & \colhead{V(14.5)}  & \colhead{FI(14.5)} & \colhead{FI(4.8)}  &  \colhead{$\alpha_{av}$} & \colhead {VLB$_{43,15}$} \\
\colhead{(1)} & \colhead{(2)} & \colhead{(3)} & \colhead{(4)}  & \colhead{(5)} &
\colhead{(6)} & \colhead{(7)} & \colhead{(8)} & \colhead{(9)}  & \colhead{(10)}
}
\startdata
0016$+$731 &          & Q  & c       & 1.781  & 0.62  & 0.38 & 0.32  & $-$0.04 & K,L \\
0040$+$517 & 3C 20    & G  & ld      & 0.174  & 0.19  & 0.06 & -0.01  & $-$1.02 &    \\
0108$+$388 & OC 314   & G  & cso     & 0.668  & 0.25  & 0.09 & 0.04  & $-$0.91 &    \\
0133$+$476 &  DA 55   & Q  &  c      & 0.859  & 0.71  & 0.39 & 0.29  & $+$0.27 & K,L  \\
0153$+$744 &          & Q  & cd      & 2.338  & 0.31  & 0.14 & 0.09  & $-$0.93 & K,L \\ 
0212$+$735 &          & Q  & a1      & 2.367  & 0.31  & 0.16 & 0.15  & $+$0.03 & K,L \\
0220$+$427 &  3C 66B  & G  & ld      & 0.021  & 0.30  & 0.14 & 0.03  & $-$1.32 &   \\
0315$+$416 & 3C 83.1B & G  & ld      & 0.025  & 0.34  & 0.13 & 0.04  & $-$1.10 &   \\
0316$+$413 &  3C 84   & G  & i       & 0.018  & 0.49  & 0.29 & 0.34  & $-$0.19 & L \\
0404$+$768 & 4C 76.03 & G  & cso     & 0.599  & 0.12  & 0.04 & 0.01  & $-$0.60 &  \\
0538$+$498 &  3C 147  & Q  & ssc     & 0.545  & 0.26  & 0.10 & 0.01  & $-$0.98 & L  \\
0605$+$480 &  3C 153  & G  & ld      & 0.277  & 0.10  & -0.02 & 0.03  & $-$1.08 &   \\
0710$+$439 &   OI 417 & G  & cso     & 0.518  & 0.11  & 0.03 & 0.03  & $-$0.64 & K  \\
0711$+$356 & OI 318   & Q  & cd      & 1.62   & 0.27  & 0.07 & 0.12  & $-$0.67 & L \\
0723$+$679 & 3C 179   & Q  & ld      & 0.846  & 0.47  & 0.14 & 0.07  & $-$0.35 & L   \\
0804$+$499 &  OJ 508  & Q  & c       & 1.43   & 0.64  & 0.30 & 0.24  & $+$0.05 & K,L,J \\
0809$+$483 & 3C 196   & Q  & ld      & 0.871  & 0.18  & 0.03 & 0.01  & $-$1.15 &   \\
0814$+$425 & OJ 425   & BL & c       & 0.245  & 0.57  & 0.34 & 0.22  & $-$0.06 & K,L   \\
0831$+$557 & 4C 55.16 & G  & i       & 0.242  & 0.24  & 0.12 & 0.00  & $-$1.11 &   \\
0836$+$710 & 4C 71.07 & Q  & a2      & 2.172  & 0.47  & 0.26 & 0.09  & $-$0.25 & K,L,J \\
0850$+$581 & 4C 58.17 & Q  & c       & 1.322  & 0.20  & 0.09 & 0.03  & $-$0.40 & K,L \\
0859$+$470 & 4C 47.29 & Q  & a1      & 1.462  & 0.26  & 0.12 & 0.09  & $-$0.29 & L \\
0906$+$430 & 3C 216   & Q  & ssc     & 0.670  & 0.22  & 0.08 & 0.04  & $-$0.40 & L \\
0917$+$458 & 3C 219   & G  & ld      & 0.174  & 0.17  & 0.04 & 0.03    & $-$1.21 &  \\
0923$+$392 & 4C 39.25 & Q  & cd      & 0.695  & 0.58  & 0.45 & 0.24  & $-$0.05 & K,L \\ 
0945$+$408 & 4C 40.24 & Q  & a2      & 1.252  & 0.32  & 0.20 & 0.21  & $-$0.15 &  K,L \\
0951$+$699 & M 82     & G  & \nodata & 0.001  & 0.12  & 0.06 & 0.01  & $-$0.68 &   \\
0954$+$556 & 4C 55.17 & Q  & \nodata & 0.909  & 0.24  & 0.10 & 0.02  & $-$0.41 &  L \\
0954$+$658 &          & BL & \nodata & 0.368  & 0.82  & 0.38 & 0.49  & $+$0.04 & L,J \\
1003$+$351 & 3C 236   & G  & ld      & 0.101  & 0.17  & 0.08 & 0.04  & $-$0.61 &  \\  
1031$+$567 & OL 553   & G  & cso     & 0.460  & 0.35  & 0.15 & 0.03  & $-$0.97 &  \\
1157$+$732 & 3C 268.1 & G  & ld      & 0.970  & 0.13  & 0.06 & 0.02  & $-$0.91 &  \\
1254$+$476 & 3C 280   & G  & ld      & 0.996  & 0.12  & 0.03 & 0.03  & $-$1.05 &  \\
1358$+$624 & 4C 62.22 & G  & cso     & 0.431  & 0.18  & 0.05 & 0.02  & $-$0.84 &  \\
1409$+$524 & 3C 295   & G  & ld      & 0.464  & 0.12  & 0.02 & 0.02  & $-$1.24 &  \\
1458$+$718 & 3C 309.1 & Q  & ssc     & 0.905  & 0.38  & 0.19 & 0.05  & $-$0.59 & K \\
1609$+$660 & 3C 330   & G  & ld      & 0.550  & 0.15  & 0.05 & 0.02  & $-$1.17 &  \\
1624$+$416 & 4C 41.32 & Q  & a2      & 2.550  & 0.30  & 0.17 & 0.11  & $-$0.39 & L \\
1633$+$382 & 4C 38.41 & Q  & a2      & 1.814  & 0.33  & 0.15 & 0.13  & $-$0.16 & K,L,J \\
1634$+$628 & 3C 343   & Q  & ssc     & 0.988  & 0.20  & 0.11 & 0.02  & $-$1.24 &  \\
1637$+$574 & OS~562   & Q  & c       & 0.751  & 0.54  & 0.27 & 0.18  & $+$0.16 & L \\
1641$+$399 & 3C 345   & Q  & a2      & 0.593  & 0.47  & 0.26 & 0.24  & $+$0.02 & K,L \\
1642$+$690 & 4C 69.21 & Q  & c       & 0.751  & 0.43  & 0.14 & 0.13  & $+$0.00 & K,L \\
1652$+$398 & Mkn~501  & BL & a2      & 0.034  & 0.20  & 0.07 & 0.09  & $-$0.25 & K,L,J \\
1739$+$522 & 4C 51.37 & Q  & c       & 1.375  & 0.71  & 0.46 & 0.51  & $+$0.00 & K,L,J \\
1749$+$701 &          & BL & a1      & 0.770  & 0.41  & 0.17 & 0.11  & $-$0.16 & K,L \\
1803$+$784 &          & BL & a1      & 0.680  & 0.31  & 0.12 & 0.13  & $+$0.03 & K,L \\
1807$+$698 & 3C~371   & BL & a2      & 0.051  & 0.32  & 0.14 & 0.09  & $-$0.12 & K,L \\
1823$+$568 & 4C 56.27 & BL & a1      & 0.664  & 0.42  & 0.21 & 0.10  & $+$0.09 & K,L \\
1828$+$487 & 3C 380   & Q  & ssc     & 0.692  & 0.18  & 0.09 & 0.05  & $-$0.56 & K,L \\
1842$+$455 & 3C 388   & G  & ld      & 0.091  & 0.14  & 0.04 & 0.02  & $-$1.02 &  \\
1845$+$797 & 3C 390.3 & G  & ld      & 0.056  & 0.19  & 0.12 & 0.04  & $-$0.94 & K \\
1928$+$738 & 4C 73.18 & Q  & a1      & 0.302  & 0.24  & 0.11 & 0.06  & $-$0.09 & K,L \\
1939$+$605 & 3C 401   & G  & ld      & 0.201  & 0.17  & 0.02 & 0.03  & $-$1.11 &  \\
1954$+$513 & OV 591   & Q  & a1      & 1.220  & 0.48  & 0.27 & 0.13  & $-$0.10 & K,L \\
2021$+$614 & OW 637   & Q  & cso?    & 0.227  & 0.22  & 0.13 & 0.05  & $-$0.09 & K,L \\
2153$+$377 & 3C 438   & G  & ld      & 0.290  & 0.19  & -0.04 & 0.03  & $-$1.30 &  \\
2200$+$420 & BL~LAC   & BL & a1      & 0.069  & 0.59  & 0.32 & 0.27  & $-$0.02 & K,L,J \\
2229$+$391 & 3C 449   & G  & ld      & 0.017  & 0.23  & 0.08 & 0.04  & $-$1.27 &  \\
2243$+$394 & 3C 452   & G  & ld      & 0.081  & 0.35  & 0.08 & 0.02   & $-$1.48 &  \\
2351$+$456 & 4C~45.51 & Q  & a2      & 1.992  & 0.46  & 0.25 & 0.16  & $-$0.17 &  L \\
2352$+$495 & DA 611   & G  & cso     & 0.237  & 0.14  & 0.05 & 0.03  & $-$0.69 &  \\
\enddata

\end{deluxetable}
\clearpage

\begin{deluxetable}{crlcrclrrrrrrrl}
\tabletypesize{\scriptsize}
\tablecaption{POLARIZATION AND SMALL-SCALE STRUCTURE ORIENTATIONS. 
\label{tbl-2}}
\tablewidth{540pt}
\rotate
\tablenum{2}
\tablecolumns{15}
\tablehead{\colhead{Source} & \colhead{RM} & \colhead{Ref} & \colhead{P\%$_{av}$(14.5)} &  \colhead {sd}
& \colhead{$\chi_{0}(14.5)$} & \colhead{\%} & \colhead{$\chi_{0}(4.8)$} &  \colhead{\% } & \colhead{$\Theta_{5}$} & \colhead{Ref} & \colhead{$\Theta_{15}$} & \colhead{Ref} & \colhead{$\Theta_{43}$} & \colhead{Ref} \\
\colhead{(1)} & \colhead{(2)} & \colhead{(3)} & \colhead{(4)}  & \colhead{(5)} &
\colhead{(6)} & \colhead{(7)} & \colhead{(8)} & \colhead{(9)}  & \colhead{(10)} &
\colhead{(11)} & \colhead{(12)} & \colhead{(13)} & \colhead{(14)}  & \colhead{(15)}
 }
\startdata
0016$+$731 & $-$3    & Ru  & 1.36 &  0.16 & np  &    &  66 & 25 & 156 & P1  &  u    & KZ  & 132 & L \\
0040$+$517 & $+$159  & SN  & 1.88 &  0.15 & 57  & 16 &  19 & 38 & wc  & P1  & \nodata  &   & \nodata  &   \\
0108$+$388 & $+$3    & Ru  & 0.44 &  0.37 & np  &    &  np &    & 237 & P1  & \nodata  &    & \nodata  &  \\
0133$+$476 & $+$28   & RJ  & 0.15 &  0.08 & np  &    &  22 & 35 & 317 & W   & 321   & KZ  & 330 &  L \\
0153$+$744 & $-$42   & Ru  & 3.82 &  0.43 & np  &    &  np &    & 156 & P1  & 101   & KZ  & 68  & L  \\ 
0212$+$735 & $+$14   & Ru  & 0.69 &  0.09 & 30  & 25 &  33 & 77 & 103 & P1  & 111   & KZ  & 121 &  L \\ 
0220$+$427 & \nodata &     & 2.79 &  0.28 & np  &    & 71  & 98 &  \nodata  &  &  \nodata   &     &  \nodata  & \\ 
0315$+$416 & $+$18   & SN  & 4.27 &  0.41 & 115 & 27 & 105 & 98 & 90  & Xu  & \nodata  &     & \nodata    &    \\
0316$+$413 & $+$76   & Ru  & 0.12 &  0.01 & 14  & 19 &  11 & 18 & 190 &  W  &  190  & HW  & 176 & L \\ 
0404$+$768 & \nodata &     & 2.79 &  0.24 & 42  & 39 &  np &    & 48  & P1  & \nodata  &     &  \nodata   &   \\  
0538$+$498 & $-$1300 & Na  & 2.77 & 0.07  & 91 & 81 & np  &     & 230 & P1  &  \nodata &     & 237 & L  \\  
0605$+$480 & $+$34 & SN    & 5.88 & 0.51  & np  &    & 45  & 95 & wc  & P1  & 112   & KZ  & \nodata & \\   
0710$+$439 & $-$7  & Ru    & 0.50 &  0.32 & np  &    & np  &    &  0  & P1  & \nodata  &     & \nodata &   \\   
0711$+$356 & $+$40 & St    & 1.09 & 0.37  & np  &    & np  &    & 338 & P1  &  \nodata &     & 329 & L  \\  
0723$+$679 & $-$25:& Ru    & 0.98 & 0.36  & np  &    & 177 & 78 & 272 & P1  & \nodata  &     & 256 &  L \\ 
0804$+$499 & $+$2  & Ru    & 1.40 & 0.13  & np  &    & 85  & 46 & 115 & XP &  130  & KZ  & 127 & L  \\  
0809$+$483 & $-$142 & SN   & 1.40 & 0.17  & np  &    & 153 & 100 & wc & P1  &  \nodata &     &  \nodata &    \\  
0814$+$425 & $+$19 & Ru    & 2.40 &  0.17 & 92  & 35 &  83 & 40 & 130 & GP  & 84b   & KZ  & 103 & L \\ 
0831$+$557 & $+$113 & Ru   & 0.49 & 0.14  & np  &    &  np &    & 292: & P1  & \nodata &     & \nodata &   \\ 
0836$+$710 & $-$11 & W1    & 4.67 & 0.15  & 106 & 76 & 105 & 100 & 214 & P1  & \nodata &     & 201 & L \\  
0850$+$581 & $-$6  & Ru    & 4.07 & 0.29  & 69  & 29   & np  &   & 156 & P1  & 176  & KZ  & 227 & L \\  
0859$+$470 & $-$31 & RJ    & 1.55 & 0.31  & np  &    & 92  & 80 &   4  & XP  & \nodata    &     & 357 & L \\  
0906$+$430 & $+$28 & RJ    & 0.82 & 0.14  & np  &    & 84  & 80 & 155c & P1  & \nodata    &     & 151 & L \\  
0917$+$458 & $-$19 & SN    & 2.86 & 0.30  & np  &    & 148 & 94 &  wc  & P1  & \nodata    &     & \nodata    & \\ 
0923$+$392 & $+$15 & RJ    & 2.73 & 0.08  & 134 & 83 & 110 & 55 &  277 & P1  & 278   & KZ  & 93  & L \\
0945$+$408 & $+$7  & RJ    & 1.34 & 0.16  & np  &    &  15 & 98 & 127c & P1  &  u    & KZ  & 137 & L \\
0951$+$699 & \nodata &     & 1.24 & 0.17  & np  &    & 171 & 38 & 240  & P1  & \nodata &     & \nodata &  \\  
0954$+$556 & $+$1  & SN    & 3.98 & 0.19  & 5   & 67 &   4 & 94 & \nodata &     & \nodata &     & 191 & L \\
0954$+$658 & $-$15 & Ru    & 4.88 & 0.24  & 166 & 39 & 175 & 72 & 306  & P1  & \nodata &     & 311 & L \\
1003$+$351 & $-$4  & Ru    & 0.79 & 0.33  & np  &    & 107 & 18 & 292  & P1  & \nodata &     & \nodata &  \\  
1031$+$567 & \nodata &     & 0.65 & 0.55  & np  &    & np  &    & \nodata &     & \nodata  &  & \nodata  &  \\
1157$+$732 & $+$8  & SN    & 5.00  &  0.38 & 115 & 43 & 143 & 79 &  wc   &    & \nodata &   & \nodata   &  \\
1254$+$476 & $-$17 & SN    & 10.99 & 0.37 & 54  & 78 &  48 & 100 & wc    &    & \nodata &   & \nodata  &  \\
1358$+$624 & \nodata  &    &  1.64 &  0.23 & np  &    &  np &    & 305:  & P1 & \nodata &   & \nodata  & \\
1409$+$524 & \nodata  &    &  1.56 &  0.10 & 105 & 21   &  np &    &  wc & P1 & \nodata &   & \nodata &  \\
1458$+$718 & $+$75:& Ru    & 2.20  & 0.23  & 50  & 38 & 46  & 98 &  164c & P1 & \nodata &   & \nodata &   \\
1609$+$660 & $+$13 & SN    & 3.88  & 0.35  & np  &    & 129 & 94 &  wc   & P1 & \nodata &   &  \nodata &  \\
1624$+$416 & $+$1  & Ru    & 1.01  & 0.27  & np  &    & np  &    & 239c  & P1 & \nodata &   & 262 &  L \\
1633$+$382 & $+$15 & RJ    & 0.24  & 0.08  & np  &    &  15 & 42 & 295   & P1  & 282   & KZ  & 279 & L \\
1634$+$628 & $+$0  & Ru    & 2.36  & 0.49  & np  &    &  55 & 20 & \nodata  &     & \nodata  & &  \nodata   &  \\
1637$+$574 & $+$22 & RJ    & 2.28  & 0.15  & 118 & 50 & 111 & 76 & 190   & XP  & \nodata &     & 200 & L \\
1641$+$399 & $+$29 & RJ    & 1.52  & 0.07  & 37  & 61 &  24 & 94 & 230c  & P1  & 270   & KZ  & 284 & L \\
1642$+$690 & $-$11 & W1    & 6.38  & 0.16  & 159 & 80 & 116 & 32 & 195   & P1  &  195b & KZ  & 158 & L \\
1652$+$398 & $+$42 & Ru    & 2.22  & 0.14  & 178 & 48 & 177 & 37 & 130   & P1  & 147b  & KZ  & 160 & L \\
1739$+$522 & $-$4  & Ru    & 1.07  & 0.13  &  np &    & np  &    &  25   & XP  & \nodata &   & 204 & L \\
1749$+$701 & $+$ 8 & W1    & 5.04  & 0.28  &  85 & 23   & 89  & 38 & 296c & P1  & 301c  & KZ  & 275 & L \\
1803$+$784 & $-$61 & W1    & 3.42  & 0.09  & 98 & 70 & 91  & 80 & 268c   & P1  & 267c  & KZ  & 291 & L \\
1807$+$698 & $+$200 & Wr   & 1.75  & 0.08  & 151 & 22 & 162  & 75 & 263  & P1  & 263   & KZ  & 252 & L \\
1823$+$568 & $+$36 & W1    & 7.30  & 0.17 & 29 & 86 & 29  & 96 & 197c    & P1  & 202c  & KZ  & 201 & L \\
1828$+$487 & $+$26:& Ru    & 0.74  & 0.07  & np &    & 30  & 98 & 328    & P1  & \nodata  &  & 311 & L \\
1842$+$455 & \nodata &     & 1.54  & 0.31  & np &    & 174 & 87 &  wc    & P1  &  \nodata &  & \nodata &   \\
1845$+$797 & $-$5    & SN  & 4.39  & 0.19  & 41 & 66 & 27  & 100 & 322   & P1  & 324   & KZ  &  \nodata &  \\
1928$+$738 & $+$33 & W1    & 2.44  & 0.06  & 79 & 79 & 79  & 98 & 166    & P1  & 163   & KZ  & 166 & L \\
1939$+$605 & \nodata  &    & 2.02  & 0.42  & np &    & 35  & 87 &  wc    & P1  & \nodata  &     & \nodata &  \\
1954$+$513 & \nodata  &    & 0.99  & 0.20  & np &    & np  &    &  294   & XP  & \nodata  &     & 307 &  L \\
2021$+$614 & $-$99 & Ru    & 0.00  & 0.09  & np &    & np  &    & 33    & P1  &  31    & KZ  & 214 & L \\
2153$+$377 & \nodata &     & 2.80  & 0.47  & np &    & np  &    & wc    & P1  & \nodata  &  &  \nodata   & \\
2200$+$420 & $-205$ & RJ   & 3.66  & 0.10  & 37 & 63 & 41  & 70 & 188   & P1 & 189   & KZ  & 209 & L \\
2229$+$391 & $-$162& SN    & 1.72  & 0.55  & np &    & 106 & 92 & wc    & P1  & \nodata  &     & \nodata &  \\
2243$+$394 & $-$274 & RZK  & 5.75  & 0.35  & 177 & 39 &  3 & 98 & 82   & P1  & \nodata   &     &  \nodata   & \\
2351$+$456 & $-$12: & Ru   & 2.27  & 0.19  & 110 & 34 & 80  & 41 & 298  & P1  & \nodata  &     & 321 & L \\
2352$+$495 & \nodata &     & 1.21  & 0.31  & np  &    & np  &    & 339  & P1  & \nodata  &     & \nodata & \\
\enddata

\tablerefs{RMs: (Na) \citet{nan00}; (RJ) \citet{rud83}; (Ru) \citet{rusk88}; 
(RZK) \citet{rzk83}; (SN) \citet{sim81}; (St) \citet{stan98}; (W1) \citet{wro93};
(Wr) \citet{wro87}.  
\newline Theta: (GP) \citet{gabp00} ; (KZ) \citet{kel98}: measured from maps at www.cv.nrao.edu/2cm survey; 
 (L) \citet{lis01a}; (HW) measured from maps in \citet{hom99}; (P1) See paper 1,
\citet{all92}; (W) \citet{weh96}; (Xu) \citet{xu99}; (XP) Xu, W. \& Polatidis, A. Private
 Communication. }
\tablecomments{RM notes. 3C~295: our data are consistent with a large, time-variable RM.
3C 371: the value adopted differs from that in \citet{all92}. A range of values exists in the literature
 for this source. We have used the value which is most consistent with our data. 
1928+738: \citet{tay00a} finds a position-dependent RM. Our frequency-dependent spread is not consistent with 
the large RM he finds for component A.
BL Lac: we use the value given in \citet{rud83}; recent VLBA observations 
suggest that this source property is time variable \citep{mut00,rey01}. 
\newline Theta notes. We adopt the value indicative of the inner flow. Most jets are kinked or
bent. b \& c flag structure showing sharp
bends or curved structure. In these cases the flow
direction is highly-position dependent.
3C~83.1: source exhibits twin jets on parsec scales.}

\end{deluxetable}

\clearpage

\begin{deluxetable}{ccccccl}
\tablenum{3}
\tablecaption{Polarization Spectra of Lobe-Dominated Objects. \label{tbl-3}}
\tablewidth{0pc}
\tablecolumns{7}
\tablehead{\colhead{Source} & \colhead{opt. class} & 
 \colhead{P(\%)$_{av(14.5)}$} &  \colhead{P(\%)$_{av(4.8)}$} & 
 \colhead{ Pol. Spectrum} &  \colhead{FR class} & \colhead{l (kpc)} \\
\colhead{(1)} & \colhead{(2)} & \colhead{(3)} & \colhead{(4)}  & \colhead{(5)} &
\colhead{(6)} & \colhead{(7)} 
 }
\startdata
0040$+$517 &  G & 1.88   &  0.50  & F/I  &  II  &  202.3 \\
0220$+$427 &  G & 2.79   &  3.59  & S    &   I  &  [90]   \\
0315$+$416 &  G & 4.27   &  5.88  & S    &   I  &  [36]   \\
0605$+$480 &  G & 5.88   &  3.97  & S/F  &  II  &  37.7    \\
0723$+$679 &  Q & 0.98   &  3.78  & F    &  II  &  124.7   \\
0809$+$483 &  Q & 1.40   &  2.32  & S    &  II  &   45.1    \\
0917$+$458 &  G & 2.86   &  3.20  & S    &  II  &  553.4   \\
1003$+$351 &  G & 0.79   &  0.70  & I    &  II  &  5853.3     \\
1157$+$732 &  G & 5.00   &  1.73  & F/I  &  II  &    365.0  \\
1254$+$476 &  G & 10.99  &  7.60  & S    &  II  &   109.8 \\
1409$+$524 &  G & 1.56   &  0.04  & I/F  &  II  &    32.3  \\
1609$+$660 &  G & 3.88   &  3.59  & S/F  &  II  &   450.2 \\
1842$+$455 &  G & 1.54   &  1.68  & S/F &  II  &  70.4   \\
1845$+$797 &  G & 4.39   &  6.40  & S    &  II  &  315.2  \\
1939$+$605 &  G & 2.02   &  2.04  & F/S    &  II  &  80.5  \\
2153$+$377 &  G & 2.80   &  0.35  & I    &  II  &   102.4 \\
2229$+$391 &  G & 1.72   &  5.82  & S    &   I  &   [38]     \\
2243$+$394 &  G & 5.75   &  7.10  & S    &  II  &  534.1  \\
\enddata

\end{deluxetable}

\clearpage

\begin{deluxetable}{ccccc}
\tablenum{4}
\tablecaption{Orientation Difference for BL Lacs. \label{tbl-4}}
\tablewidth{0pc}
\tablecolumns{5}
\tablehead{\colhead{Source} & \colhead{$\vert \chi_{0}(4.8)-\theta_{15}\vert$}
&  \colhead{$\vert\chi_{0}(4.8)-\theta_{43}\vert$}
 &  \colhead{$\vert \chi_{0}(14.5)-\theta_{15}\vert$} & \colhead{$\vert \chi_{0}(14.5)-\theta_{43}\vert$} 
 }
\startdata
0814$+$425 & 1       & 20  & 8        & 11  \\
0954$+$658 & \nodata & 44  & \nodata  & 35  \\
1652$+$398 & 30      & 17  & 31       & 18  \\
1749$+$701 & 32      & 6   & 36       & 10  \\
1803$+$784 &  4      & 20  & 11       & 13    \\
1807$+$698 & 79      & 90  & 68       & 79   \\
1823$+$568 & 7       &  8  &  7       & 8   \\
2200$+$420 & 32      & 12  & 28       & 8    \\
\enddata

\end{deluxetable}

\end{document}